\documentclass[conference]{IEEEtran}
\IEEEoverridecommandlockouts

\usepackage{amsmath,amssymb,amsfonts}
\usepackage{algorithmicx}
\usepackage{algorithm}
\usepackage{algpseudocode}
\usepackage{graphicx}
\usepackage{subcaption}
\usepackage{textcomp}
\usepackage{xcolor}
\usepackage{adjustbox}
\usepackage{enumitem}
\usepackage{amsmath}
\usepackage{amssymb}
\usepackage{MnSymbol}
\usepackage{adjustbox}
\usepackage{varwidth}
\usepackage[normalem]{ulem}
\usepackage[table]{xcolor}
\usepackage{booktabs}
\usepackage{multirow}
\usepackage{makecell}
\usepackage{mwe}
\usepackage{xurl}

\newcolumntype{P}[1]{>{\centering\arraybackslash}p{#1}}

\usepackage[
  backend=biber,
  style=ieee,
  citestyle=numeric-comp,
  giveninits=true,
  maxbibnames=3,
  minbibnames=1,
  maxcitenames=2,
  mincitenames=1,
  doi=false,
  url=false,
  isbn=false,
  eprint=false
]{biblatex}

\DeclareSourcemap{
  \maps[datatype=bibtex]{
    \map{
      \step[fieldsource=booktitle, match=\regexp{International}, replace={Int.}]
      \step[fieldsource=journal, match=\regexp{International}, replace={Int.}]
      \step[fieldsource=booktitle, match=\regexp{Conference}, replace={Conf.}]
      \step[fieldsource=booktitle, match=\regexp{Proceedings}, replace={Proc.}]
    }
  }
}

\AtEveryBibitem{%
  \clearlist{publisher}%
  \clearfield{publisher}%
  \clearlist{location}%
  \clearfield{location}%
}

\usepackage[normalem]{ulem}

\newcommand{\replacenewa}[2]{%
    {\color{black}#2}%
}

\newcommand{\replaceieeecr}[2]{%
  {\color{red}}%
    {\color{black}#2}%
}

\newcommand{\replaceieee}[2]{%
    {\color{black}#2}%
}

\renewcommand{\times}{\cdot}

\newtheorem{definition}{Definition}
\newtheorem{observation}{Key Observation}
\newtheorem{example}{Example}

\newcommand{\maybe}[1]{}

\newcommand{\nsbp}{\nobreak\kern\fontdimen2\font\relax}

\def\diaend{\hspace*{\fill} $\diamond$}

\def\clubend{\hspace*{\fill} $\clubsuit$}

\def\BibTeX{{\rm B\kern-.05em{\sc i\kern-.025em b}\kern-.08em
    T\kern-.1667em\lower.7ex\hbox{E}\kern-.125emX}}

\begin{document}

\title{VOS: Variate Ordering Strategies for Skyline Query Optimization}

\author{\IEEEauthorblockN{Abhinav Gorantla}
\IEEEauthorblockA{
\textit{Arizona State University}\\
Tempe, USA\\
agorantla@asu.edu}
\and
\IEEEauthorblockN{Pratanu Mandal}
\IEEEauthorblockA{
\textit{Arizona State University}\\
Tempe, USA\\
pmandal5@asu.edu}
\and
\IEEEauthorblockN{K. Sel\c{c}uk Candan}
\IEEEauthorblockA{
\textit{Arizona State University}\\
Tempe, USA\\
candan@asu.edu}
\and
\IEEEauthorblockN{Maria Luisa Sapino}
\IEEEauthorblockA{
\textit{University of Turin}\\
Turin, Italy\\
mlsapino@di.unito.it}
\thanks{The work  was supported in part by NSF grant \#2311716, 
\replaceieeecr{"}{``}CausalBench: A Cyberinfrastructure for Causal-Learning Benchmarking for Efficacy, Reproducibility, and Scientific Collaboration\replaceieeecr{"}{''}. }
}







\maketitle

\begin{abstract}

Efficiency of skyline algorithms is highly influenced by the underlying data characteristics. Traditionally, optimization efforts have focused on minimizing the total number of tuple-pair dominance checks to improve query performance. However, in practice, a dominance check between two tuples does not necessarily require evaluating dominance relationships for each and every preference attribute of the data
and this creates a disconnect between dominance checks optimization and query execution performance.
In this paper, we argue that skyline algorithms need to optimize total per-attribute dominance checks, along with per-tuple dominance checks and that, for both of these goals, 
the ordering of the attributes (or variates) can have a substantial impact on the efficiency of skyline computation. 
Based on this premise, we present
several strategies for identifying an  effective variate order to minimize redundant attribute comparisons. Extensive experiments on both synthetic and real-world datasets, and on both scalar and SIMD architectures, confirm the effectiveness of the proposed approach in reducing computational overhead and improving skyline query performance.

\end{abstract}


\section{Introduction}\label{sec:introduction}

Skyline queries are an important tool in multi-criteria decision support%
\replaceieeecr{}{~\cite{ZHANG2026130889, 7373349, needle_in_haystack}}.
They aim to extract the Pareto-optimal set of tuples given a set of per-attribute preference criteria\replaceieeecr{}{~\cite{Borzsonyi01theskyline}}, offering key insights into the data set and its distribution. More specifically, given a set $D$ of data tuples in a variate/attribute\footnote{We use the terms {feature,} variate, and attribute interchangeably.} space $A$, a set of preference attributes $P \subseteq A$, and a set of per-attribute min/max preference criteria, $\Theta$, the skyline $S$ of $D$ consists of a subset of tuples that are not dominated by any other tuple in the dataset $D$. A data point dominates another data point if it is as good or better in all preference attributes, and strictly better in at least one attribute. We define this formally in Section \ref{sec:prelims:domDefns}.

\begin{table}[t]
    \centering
    \caption{
    Running example: data for hyperparameter selection problem; highlighted tuples belong to the skyline\protect\footnotemark}\label{tab:hyperparameter-data}
    \scriptsize
    \vspace*{-0.05in}
    \begin{adjustbox}{width=\columnwidth}
        \begin{tabular}{|c|c|c|c|c|}
        \hline
        Tuple & HP\_Conf & MSE\_Loss & Huber\_Loss & Model\_Size (MB) \\
        \hline\hline
        $t_0$ &  $\vec{h_0} = \langle\cdots\rangle$   & 0.4     & 0.2     & 957 \\
        \rowcolor[rgb]{1, 1, 0}
        $t_1$ &  $\vec{h_1} = \langle\cdots\rangle$  & 0.1    & 0.3     & 228 \\
        $t_2$ &  $\vec{h_2}  = \langle\cdots\rangle$  & 0.6     & 0.2     & 613 \\
        \rowcolor[rgb]{1, 1, 0}
        $t_3$ &  $\vec{h_3} = \langle\cdots\rangle$  & 0.4     & 0.1     & 319 \\
        \rowcolor[rgb]{1, 1, 0}
        $t_4$ &  $\vec{h_4} = \langle\cdots\rangle$  & 0.6     & 0.3 & 183 \\
        $t_5$ &  $\vec{h_5} = \langle\cdots\rangle$  & 0.9     & 0.4     & 563 \\
        \rowcolor[rgb]{1, 1, 0}
        $t_6$ &  $\vec{h_6} = \langle\cdots\rangle$  & 0.8     & 0.2     & 284 \\
        $t_7$ & $\vec{h_7} = \langle\cdots\rangle$   & 0.8     & 0.3     & 603 \\
        \rowcolor[rgb]{1, 1, 0}
        $t_8$ &  $\vec{h_8} = \langle\cdots\rangle$  & 0.3     & 0.1     & 809 \\
        $t_9$ &  $\vec{h_9} = \langle\cdots\rangle$  & 0.6     & 0.1     & 407 \\
        \hline
        \end{tabular}
    \end{adjustbox}
\end{table}
\footnotetext{
Due to an editorial oversight, the $t_1$ row was not highlighted as a skyline entry in the previous version of this manuscript~\cite{gorantla2026vosvariateorderingstrategies}; this error has been corrected in this version.}
\begin{example}[Hyperparameter selection]
\label{ex:hyperparameter}
\phantom{.}
\replaceieee{}{Let us consider 
a machine learning researcher who aims to 
pick among grid search runs a hyperparameter configuration, \texttt{HP\_Conf},  that leads to small \texttt{MSE\_Loss}, \texttt{Huber\_Loss}, and \texttt{Model\_Size}~\cite{causalbench_er,causalbenchgrant}. This task can be formulated as a search for the non-dominated set of configurations with respect to the preference attributes \mbox{$P = \{\texttt{MSE\_Loss},\, \texttt{Huber\_Loss},\, \texttt{Model\_Size}\}$} and {\em mini-mization} preference criterion.
%
Table~\ref{tab:hyperparameter-data} shows a sample data set and the corresponding skyline  based on these preference criteria. In this example, the skyline  provides alternative configurations for the researcher to consider.}
\clubend \end{example}


\subsection{Tuple-Pair  vs. Per-Attribute Dominance Checks}\label{sec:argument_1}

\begin{figure}[t]
\centering{%
   \begin{tabular}{cc}
    \includegraphics[width=0.45\columnwidth]
    {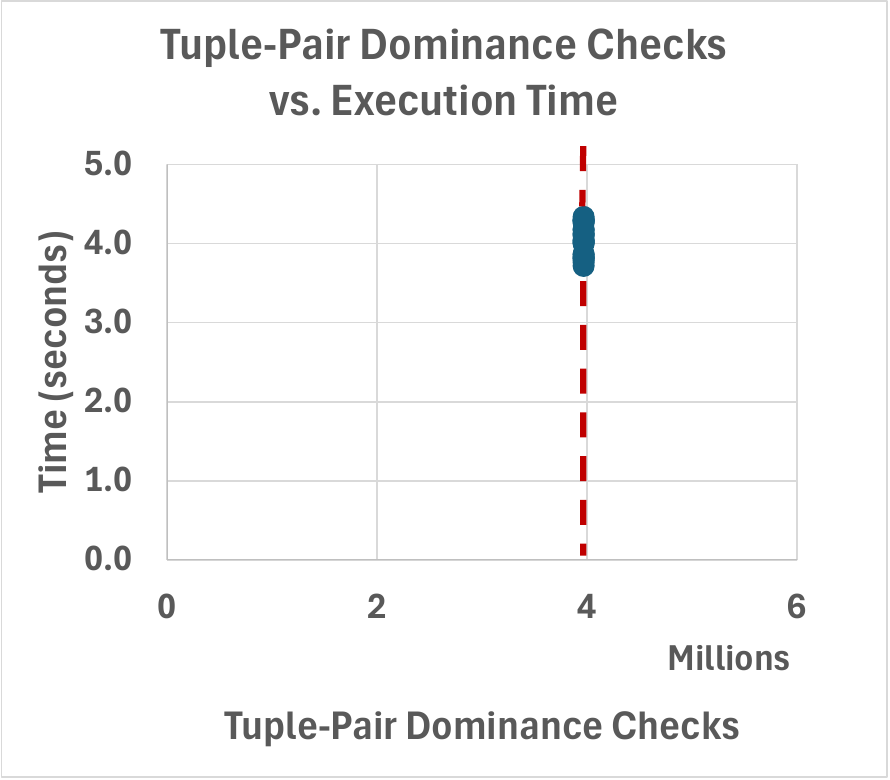}
    &
    \includegraphics[width=0.45\columnwidth]
    {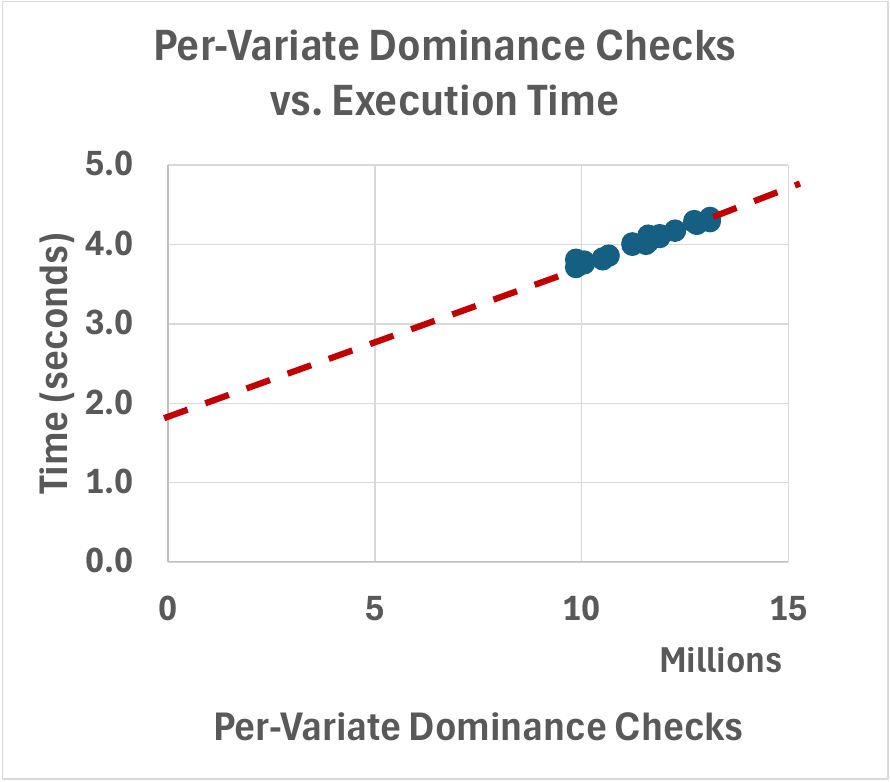}
    \\
    \makecell{(a) Tuple-pair\\dominance checks}
    &
    \makecell{(b) Per-variate\\dominance checks}
   \end{tabular}
}
\caption{(a) Tuple-pair dominance checks and (b) per-attribute (or per-variate) dominance checks as proxies for skyline query execution time: while the number of tuple-pair dominance checks \replaceieee{stay}{stays} constant when the order of attributes \replaceieee{change}{changes}, the number of per-variate dominance checks correlates well with the BNL skyline computation time
}
\label{fig:problem}
\vspace*{-0.05in}
\end{figure}

In Section \ref{sec:related-works}, we provide an overview of the state-of-the-art in skyline algorithms. 
Importantly, the literature has focused on reducing the \textit{total number of tuple-pair dominance checks} through pre-sorting~\cite{Chomicki03skylinewith,Salsa} and indexing~\cite{PapadiasTFS05,Zorder}.
%
%
%
%
The problem with this approach, however, is that, as we visualize in Figure~\ref{fig:problem}, the total number of tuple-pair dominance checks does not necessarily correlate well with the execution time of the skyline algorithms. 
The main reason for this is that  a dominance check between two tuples does not necessarily require evaluating dominance relationships for each and every preference attribute of the data
and this creates a disconnect between dominance check optimization and query execution performance.
Indeed, as we see in Figure~\ref{fig:problem}, the number of per-attribute dominance checks does correlate with the skyline query performance.
Therefore, in this paper, we first argue that the correct proxy measure for skyline query performance should not be the total number of tuple-pair dominance checks, but the \textit{number of per-attribute dominance checks} (or per-chunk dominance checks in SIMD architectures, where a limited degree of parallelism is possible during  evaluation). 

\subsection{Variate Orders and Per-Attribute Comparisons}

In this paper, our second argument is that, when considering the total number of per-attribute dominance checks, we also need to carefully consider the variate\replaceieee{/attribute}{} ordering, in addition to seeking suitable tuple orderings.
%
%
%
%
Let us reconsider 
Example \replaceieee{\ref{ex:flight}}{\ref{ex:hyperparameter}}:

\begin{example}{Impact of the Variate Order.}
\replaceieee{}{Let us assume that 
we are trying to determine if \replaceieeecr{}{$t_4 = (0.6,0.3,183)$} in Table~\ref{tab:hyperparameter-data} is dominated by tuple $t_3 = (0.4,0.1,319)$. If we follow the default order  $\langle\texttt{MSE\_Loss},\, \texttt{Huber\_Loss},\, \texttt{Model\_Size}\rangle$ of the attributes, all three preference attributes must be compared to conclude that $t_3$ and $t_4$ have a  non-dominating relationship. However, 
if we use a different variate order, $\langle\texttt{Model\_Size},\, \texttt{MSE\_Loss},\, \texttt{Huber\_Loss}\rangle$, this  check would be completed after comparing only the first two attributes.}
\clubend \end{example}


\begin{table}[t]
    \centering
    \caption{\replaceieee{}{\replaceieeecr{}{Correlation matrix for preference attributes in Ex.~\ref{ex:hyperparameter}}}}\label{tab:flights-correlation}
    \vspace*{-0.05in}
    \begin{adjustbox}{width=0.9\columnwidth}
        \begin{tabular}{|l|rrr|}
        \hline
        & MSE\_Loss & Huber\_Loss & Model\_Size (MB) \\
        \hline
        MSE\_Loss &       & \cellcolor[rgb]{ .388,  .745,  .482}0.39 & \cellcolor[rgb]{ .988,  .988,  1}-0.03 \\
        Huber\_Loss & \cellcolor[rgb]{ .388,  .745,  .482}0.39 &       & \cellcolor[rgb]{ .973,  .412,  .42}-0.16 \\
        Model\_Size (MB) & \cellcolor[rgb]{ .988,  .988,  1}-0.03 & \cellcolor[rgb]{ .973,  .412,  .42}-0.16 &  \\
        \hline
        \end{tabular}
    \end{adjustbox}
    \vspace*{-0.05in}
\end{table}



The above example, of course, (a)  does not establish if there is a common attribute/variate ordering that would reduce the total number of per-attribute dominance checks and (b), if such an ordering exists, it does not tell us how to identify this order.
In this paper, our answer to (a) is in affirmative and, for (b), we argue that the distribution of the data, in particular the attribute correlation matrix (Table~\ref{tab:flights-correlation}), provides us the necessary signals  to help seek a suitable variate order. 

\subsection{%
Variate Ordering and the Number of Tuple-Pair Dominance Checks}

While reducing the number of per-attribute checks is useful, this should not come at the cost of an increased number of tuple-pair comparisons.
Importantly, the order in which variates are compared during dominance check does not directly impact the number of tuple-pair comparisons {\bf unless} the variate-order is reflected explicitly in the skyline tuple enumeration process.
In this paper, we further argue that the variate ordering strategy can  have a direct and significant impact on the filtering efficiency of the  tuple ordering strategy, especially if the ordering strategy itself is variate order sensitive, as in lexicographical sort.
In fact, as we see in Section~\ref{sec:variate_order_sfs_salsa}, such a  strategy can perform better than a lexicographic order without considering the variate orders, as well as more conventional pre-sorting based techniques, such as sort-first skylines, SFS~\cite{Chomicki03skylinewith}, SaLSa~\cite{Salsa}.

\subsection{Our Contributions: Variate Ordering for Skyline Query  Optimization (VOS)}


As outlined above, in this paper, we argue that the efficiency of skyline algorithms can be significantly improved by selecting an appropriate ordering of the preference variates. We investigate the properties of various skyline algorithms and analyze how the variate ordering (and, by extension, the underlying data distribution) influences their efficiency.
 In particular, we show that

\begin{itemize}[leftmargin=*]
\item it is possible to determine whether a tuple dominates another without comparing all variates, by prioritizing comparisons on the most conflicting preference attributes first,
\vspace{0.5em}
\item the correlations\footnote{\replaceieeecr{}{In this paper, we consider Pearson correlation.}} present in the data can be  leveraged to construct strategies that approximate the optimal variate ordering for skyline computation to reduce {\bf both} the number of {\em tuple-pair dominance checks} and {\em per-attribute comparisons}.

\end{itemize}

Finally, we provide a detailed experimental analysis on the impact of variate ordering on skyline algorithms on varying sizes and distributions of synthetic data along with  real-world datasets 
\cite{Hindy_2021, uci_ml_repo}, on both scalar and SIMD  architectures.

\section{Related Works}\label{sec:related-works}
The concept of skyline queries was first introduced by Kung {\em et al.} in \cite{KungLP75} as a problem of finding the maximal set of vectors and the term \replaceieeecr{"}{``}Skyline\replaceieeecr{"}{''} was first introduced by Börzsönyi {\em et al.} in \cite{Borzsonyi01theskyline}. 
The authors proposed a Block Nested Loop (BNL)  Skyline Algorithm.   \cite{Borzsonyi01theskyline} also proposes a divide and conquer approach to skyline computation.

\noindent\textbf{Tuple Ordering by Sorting.} Following this, a pre-sorting based skyline algorithm called the Sort-Filter-Skyline (SFS) algorithm was proposed \cite{Chomicki03skylinewith}. This algorithm pre-sorts the data based on a monotonic function before performing the skyline operation. In similar lines, another algorithm Sort and Limit Skyline algorithm (SaLSa) \cite{Salsa} was proposed. This approach also pre-sorted the data on which the skyline was to be computed but in SaLSa, the authors used the same idea of pre-sorting input data to limit the number of tuples read and compared in the process of skyline computation. Kung's divide and conquer was improved upon by Bentley {\em et al.} in \cite{bentley1993fast}. They introduced the Fast Linear Expected Time (FLET) algorithm which probabilistically eliminates points that cannot be part of the skyline. 
\cite{godfrey2005maximal} showed that Kung's algorithm was inefficient with increasing dimensionality of the problem; the authors proposed Linear Elimination Sort for Skyline (LESS) algorithm which combined parts of BNL, SFS and FLET.

\noindent\textbf{Tuple Ordering by Indexing.}
An R-Tree index-structure based algorithm, Branch and Bound Skyline~\cite{PapadiasTFS05} was proposed which used the R-Tree index to efficiently prune data-points in the process of skyline computation. The authors did this by checking if the bounding box of an R-Tree node is dominated by any of the points currently in the skyline. Another index based method was introduced by Lee {\em et al.} \cite{Zorder}. They proposed an index structure called the \textit{ZBTree} which indexes and stores data based on the Z-Order curve. Parallel variants of these index-based approaches were proposed in \cite{distributedZorder, park2009parallel}, which leveraged the modern multi-core CPU architecture to optimize skyline computation.

More recently, in \cite{10.1145/3802026} 
the authors use the concepts of d-separation in causal graphs to selectively de-correlate the preference vectors, to eliminate harmful negative correlations in the data through conditioning/GroupBy operations.
%
%
%
 Other skyline optimization methods included parallelization of existing algorithms \cite{distributedZorder, park2009parallel, Wu06parallelizingskyline, StojmenovicM88}, using index structures \cite{Zorder, PapadiasTFS05, skyline_on_massive_data}. Work was also done in online computation of skylines \cite{Kossmann02shootingstars}, and estimating cardinality of skylines~\cite{luo2012sampling, 10.1007/978-3-540-24627-5_7, 1129924, 10.1145/1559845.1559899, 10.1145/3588958, bentley1978average}.
A critical advance in skyline computation has been the use of SIMD architectures for reducing the cost of implementing dominance checks through the vectorization of the dominance check operation~\cite{10.1145/1893173.1893176}. This enables comparing multiple dimensions (preference variates) in parallel and thereby significantly cut down skyline computation costs, especially in high dimensional data sets. We explore the application of the proposed variate-ordering strategies to SIMD architectures in Section~\ref{Sec:simd}.

\section{Preliminaries}\label{sec:prelim}

This section 
introduces formalisms \replaceieee{}{that} we \replaceieee{utilize throughout  the paper}{use in this paper}. 
\subsection{Tuple Dominance and Skylines}\label{sec:prelims:domDefns}

Let $D$ be a set of data tuples in a feature/attribute space $A$, and $P \subseteq A$ be a set of preference attributes. Further, let $\Theta = \{\theta_p \mid p \in P\}$ be the preference criteria related to the preference attributes $P$.

\begin{definition}[Preference Criterion]
A preference criterion $\theta_p$ for a preference attribute $p$ specifies whether it is preferred that $p$ is minimized or maximized.
\diaend \end{definition}


\begin{definition}[Dominance Relation]
Given tuples $t$ and $u$, and a preference criterion $\theta_p$, \textit{weak} dominance relation ($\succeq_{\theta_p}$) states that $t$ is as good or better than $u$ for attribute $p$.
%
Similarly, \textit{strict} dominance relation ($\succ_{\theta_p}$) states that $t$ is strictly better than $u$ for attribute $p$.
\diaend \end{definition}

\begin{definition}[Per-Variate Dominance Check]\label{def:vCheck}
    A \emph{per-variate dominance check} ({\em v-check}) is the 
    atomic 
    comparison of a single preference attribute value across two tuples, i.e., comparing $t[p]$ and $u[p]$ for some $p \in P$, for
    weak or strict 
    dominance.
\diaend \end{definition}

Given the above, we can define the dominance relationship for a given pair of tuples in $D$ as follows~\cite{Borzsonyi01theskyline}:
\vspace{\baselineskip}
\begin{definition}[Tuple Dominance]\label{def:tuple-dominance}
A tuple $t$ dominates $u$ ($t \succ u$) iff $t$ weakly dominates $u$ for all preference attributes, and strictly dominates for at least one attribute:
\begin{equation}\label{eq:tuple-dominance}
t \succ u \;\equiv\;
(\forall_{p \in P} \; t[p] \succeq_{\theta_p} u[p]) \wedge (\exists_{p \in P} \; t[p] \succ_{\theta_p} u[p])
\end{equation}
\diaend \end{definition}

\begin{definition}[Tuple-Pair Dominance Check]\label{def:dominance-check}
Given two tuples, $t$ and $u$, a {\em tuple-pair dominance check} (or {\em t-check}) is a procedure  that returns \texttt{true} if and only if $t \succ u$ holds according to Eq. \eqref{eq:tuple-dominance} and returns \texttt{false} otherwise.
\diaend \end{definition}

Finally, following~\cite{Borzsonyi01theskyline}, we formalize the concept of skyline for a given dataset as follows.

\begin{definition}[Skyline]
Given data set $D$, the skyline, $S\subseteq D$, with respect to the preference attribute set $P$ and preference criteria $\Theta$ is the \underline{maximal} subset of $D$, where
$S = \{t \in D \mid \nexists_{u \in D} \; u \succ t\}.$
\diaend \end{definition}


\subsection{Problem Formulation}\label{sec:problem}

Let $\mathcal{O} = \langle T_{order}, V_{order} \rangle$ be an {\em ordering strategy}, where  $T_{order}$ denotes the order in which tuples are read and the order  $V_{order}$ is the order in which the variates are evaluated within a tuple for per-variate dominance relationships.
Given an ordering strategy, $\mathcal{O}$, and a skyline algorithm, $\mathcal{A}$,
let $C_{\mathcal{O},\mathcal{A}}(D,\Theta)$ be  the sequence of tuple-pair dominance checks required to compute the skyline of the data set, $D$ for the preference criteria $\Theta$.
%
Let us further denote the {\em total number of per-variate dominance checks} ({\em v-checks}) required to compute the skyline of data set, $D$ for preference criteria, $\Theta$,  for the ordering strategy, $\mathcal{O}$ and algorithm $\mathcal{A}$, by $\gamma(\mathcal{O},\mathcal{A},D,\Theta)$:
\[
\gamma(\mathcal{O},\mathcal{A},D,\Theta) = \sum_{{\tt t\text{-}check}_i \in C_{\mathcal{O},\mathcal{A}}(D,\Theta)} \phi({\tt t\text{-}check}_i)
\]
where $1 \leq \phi({\tt t\text{-}check}_i) \leq |P|$ denotes the number of per-variate dominance checks (Section~\ref{sec:prelims:domDefns}) required to evaluate the $i$-th tuple-pair dominance check in the sequence, 
$C_{\mathcal{O},\mathcal{A}}(D,\Theta)$. 

Given a data set $D$ and algorithm $\mathcal{A}$, our goal is to identify an ordering strategy $\mathcal{O} = \langle T_{order}, V_{order} \rangle$ which would minimize the corresponding number of per-variate dominance checks, $\gamma(\mathcal{O},\mathcal{A},D,\Theta)$, when evaluating the skyline operator.

\section{Variate Ordering Strategies for Skyline \replaceieeecr{}{Query }Optimization (VOS)}\label{sec:variate-ordering}


%
In this section, we first investigate the properties of  skyline algorithms and how ordering strategies
influence their costs.
 
\subsection{Skylines Algorithms and Ordering Strategies}
%

\subsubsection{Skyline Algorithms that Do Not Modify the Data}\label{sec:nomod}
    For skyline algorithms, such as BNL~\cite{Borzsonyi01theskyline}, that do not modify the tuple order, $T_{order}$, tuples are processed in the order in which they are stored in the data set, $D$. However, the number of {\em v-checks} can still be influenced by the variate order, $V_{order}$. This is due to the potential early stopping during tuple-pair dominance checks.

\begin{observation}[Early Stopping Condition]\label{obs:early}
Remember from Definition~\ref{def:tuple-dominance} that, given a tuple pair $t$ and $u$, $t$ dominates $u$ under the following condition:
\[
t \succ u \;\equiv\;
\underbrace{(\forall_{p \in P} \; t[p] \succeq_{\theta_p} u[p])}_{term 1} \wedge \underbrace{(\exists_{p \in P} \; t[p] \succ_{\theta_p} u[p])}_{term2}
\]
To confirm that tuple $t$ dominates $u$, both the terms need to be satisfied. This requires comparisons across all attributes, and cannot be optimized. 
However, \underline{disproving} that  $t$ dominates $u$, may be optimized under the following conditions:
\begin{itemize}[leftmargin=*]
\item $term1$ guarantees that all attributes of tuple $t$ weakly dominate $u$. Note that, this condition can be disproven without comparing all preference attributes, as soon as we discover that $t \nsucceq_{\theta_p} u$ for any preference attribute $p$.
\vspace{0.5em}
\item $term2$ guarantees that at least one attribute of  $t$ strictly dominates $u$. This condition cannot be disproven without comparing all preference attributes.
Therefore, we should terminate the  check as soon as $term1$ is disproven. 
\end{itemize}
\end{observation}
Indeed, this early stopping condition has been widely used in most standard skyline implementations; what is missing in the literature is an investigation of the impact of the order, $V_{order}$, of the preference attributes on the skyline query performance.

 \subsubsection{Pre-Sorting based Skyline Algorithms}
 
    For pre-sorting-based skyline algorithms, such as SFS~\cite{Chomicki03skylinewith} and SaLSa~\cite{Salsa}, the data are first sorted using a monotonic function.
    A common choice is a lexicographic ordering of the 
    dataset, which depends on the variate order, $V_{order}$, and in turn determines the tuple order, $T_{order}$. For these algorithms, the total number of {\em v-checks} needed to compute the skyline data set is influenced by both $T_{order}$ and $V_{order}$.
    %
%

 \subsubsection{Data Partitioning based Skyline Algorithms}\label{sec:dps}
  
    For skyline algorithms that rely on partitioning the data into subspaces, such as D\&C~\cite{KungLP75} and BBS~\cite{PapadiasTFS05}, tuples are initially read in the order in which they are stored in the data set, $D$. However, the partitioning process depends on the variate order, $V_{order}$, which ultimately affects the tuple order, $T_{order}$. As a result, once again, the number of {\em v-checks} is influenced by both $V_{order}$ and  $T_{order}$.


\subsection{Optimal Variate Ordering}

Let us define the set that contains all possible permutations of the preference attribute set $P$ as follows:
$
\Pi_{P} = \{\pi_{1}, \ldots, \pi_{k}, \ldots, \pi_{m!}\},
$
\noindent where,
$\pi_k \in \Pi_{P}$
is a unique permutation of the variates in the preference set $P$, and $m = |P|$ is the number of variates in $P$. In the rest of this paper, we shall use $\pi_k \in \Pi_P$ to denote any one permutation of the preference attribute set $P$, such that $\pi_k = \{\pi_k^i\}_{i=1}^{m}$, where $\pi_k^i$ is the $i$-th preference attribute in the $k$-th permutation of the preference attribute set.

Given a data set $D$, a set of preference attributes $P$, and a set of
preference criteria $\Theta$, there exists an ordering of the attributes
$\mathcal{\pi} \in \Pi_{P}$
that minimizes the number of {\em v-checks} required to compute the skyline. We denote such an optimal ordering by
$\mathcal{\pi}^{*}$.
%
%
Our goal is, therefore,  to approximate this optimal ordering of preference variates $\widetilde{\pi}$, such that\footnote{Here, we simplify the notation and use $\gamma(\widetilde{\pi})$ to denote $\gamma(\mathcal{O},\mathcal{A},D,\Theta)$, where $\mathcal{O} = \langle T_{order},V_{order}\rangle$, such that $V_{order} = \widetilde{\pi}$.

} $\gamma(\widetilde{\pi}) \simeq  \gamma(\pi^*)$. 

\subsection{Number of Per-Attribute Comparisons }\label{sec:correlations_and_dominance_checks}

%
%
Consider a three-attribute setting with attributes $X$, $Y$, and $Z$. Suppose that $X$ and $Y$ are perfectly correlated, while $Z$ is perfectly anti-correlated with both. Let us further assume that we have a min-min-min (or max-max-max) as preference criteria.
Let us be given  two tuples $t$ and $u$, where  $t \nsucc u$.
\begin{itemize}[leftmargin=*]
    \item If we use the variate order $\langle X, Y, Z \rangle$ when performing the tuple-dominance check,  since $X$ and $Y$ are perfectly correlated, the comparison of the second attribute $Y$ is entirely redundant once  $X$ has been examined. 
    Nevertheless, to \emph{disprove} that $t$ dominates $u$, we must still evaluate dominance for the third variate $Z$.
    \item If we instead use the variate order $\langle Z, X, Y \rangle$,  
    after  the first attribute $Z$ is considered, the second attribute $X$ will not be redundant; yet, since the third attribute $Y$ is perfectly correlated with the second attribute $X$,  once the outcome has been determined from $Z$ and $X$, inspecting $Y$ cannot alter the result 
    and thus becomes unnecessary.
\end{itemize}
%



\label{sec:heuristics}
As we  see in the above example, the pairwise correlations among the attributes\footnote{In this paper, our approach towards optimizing skylines relies on the assumption that we are able to approximate the optimal variate order based on the correlation of the data. However, when the size of the preference attribute set is $2$, it is not possible to determine the optimal order of the variates based on the correlations alone, since the correlation matrix is symmetric. This scenario is beyond the scope of this paper.} play \replaceieee{an important}{a vital} role in \replaceieee{the optimization of}{optimizing} the number of per-variate dominance checks. 
In the rest of this section, 
we explore correlation-aware variate ordering strategies.

\subsubsection{Minimum Absolute Correlation First%
}\label{sec:NACF}

%
The first strategy, minimum absolute correlation first ({\em min-abs-correlation-first}) prioritizes variates that have low absolute correlation (i.e., close to 0) with the rest of the variates; that is, 
it selects a permutation $\pi_k$, such that
\[
\forall_{\pi_k^i,\; \pi_k^j}\;\; (i < j)\; \rightarrow\; \left(\sum_{h \neq i} |corr(\pi_k^h, \pi_k^i)| < \sum_{h \neq j} |corr(\pi_k^h, \pi_k^j)|\right).
\]

\noindent 



Intuitively, the selected order of variates places a variate that is statistically most independent from the rest of the variates early in the variate order. This implies that, if and when an early stopping condition is reached (Observation~\ref{obs:early}, Section~\ref{sec:nomod}), any per-variate dominance checks made for these early variates will not be redundant. Instead, if the early variates were allowed to have high correlations with the rest of the variates, we would also increase the likelihood of redundant per-variate dominance checks as they would be difficult to avoid.

We  verify the risks of placing variates with high  overall correlation earlier in the order (as in a {\em max-correlation-first} 
strategy) in the experiments section (Section~\ref{sec:exp}) -- we see that such a strategy increases redundant work significantly and its performance becomes close to that of the worst possible variate order.
%
Indeed, as we see next, it may be more advantageous to consider variates with overall minimum correlation (even if they are negative) early in the variate order.


\subsubsection{Minimum Correlation First}
\label{sec:NCF}
The second strategy, {\em min-correlation-first}, prioritizes comparison on preference variates which are most anti-correlated with all the other 
preference variates: to generate the variate order, we consider a permutation 
$\pi_k$ of the preference variates, such that
\[
\forall_{\pi_k^i,\; \pi_k^j}\;\; (i < j)\; \rightarrow\; \left(\sum_{h \neq i} corr(\pi_k^h, \pi_k^i) < \sum_{h \neq j} corr(\pi_k^h, \pi_k^j)\right).
\]

\noindent While at initial glance it appears counter-intuitive, this strategy has two potential advantages:   
\begin{itemize}[leftmargin=*]
\item Intuitively, this ordering furthers the idea underlying the {\em min-abs-correlation-first} strategy; if a variate is negatively correlated with the rest of the variates, \replaceieee{the corresponding}{its} per-variate evaluation is not likely to be redundant if and when there is an early stop during a tuple-pair dominance check. 
\item Moreover, for pre-sorting-based skyline algorithms, ordering the negatively correlated variates first also highlights the main Pareto trade-offs within the data, and therefore tuples that are more likely to dominate others are \replaceieee{encountered}{visited} earlier, thereby further reducing the number of dominance checks.
\end{itemize}

\noindent Indeed, as we see in the experiments (Section~\ref{sec:exp}) it tends to perform better than the de-correlation promoting approach.


\subsubsection{Incremental Minimum Correlation First}\label{sec:GNCF}
The above strategies evaluate the minimum (possibly absolute) correlation of a variate with respect to all other variates {\em globally}. However, this global perspective ignores the {\em directionality} of the dominance check process.
As an alternative, we propose an incremental strategy ({\em inc-min-correlation-first}) that 
orders the variates by selecting, at each step, the variate with the minimum correlation to the {\em remaining} variates. The intuition here is that, when we have already considered a variate, the next variate to be considered should be most conflicting with the variates that have {\em not yet been considered; those that have already been considered are not relevant in this context}. 

\subsubsection{Incremental Maximum Correlation Last}\label{sec:GXCL}
While both the {\em min-correlation-first} and {\em inc-min-correlation-first} strategies prioritize placing the most conflicting variates early in the variate order, an alternative perspective could be to place least conflicting (most correlated) variables late in the order, thereby minimizing the number of comparisons involving highly correlated variates. To this end, we propose an alternative strategy that constructs the variate order by placing variates with the highest correlations to the remaining variates last. The intuition is that variates that are strongly correlated with others contribute the least additional discriminatory power: in the extreme case when two variates are perfectly correlated, comparing one variate renders comparison on the other variate redundant. Consequently, such less informative attributes should be pushed toward the end of the variate order.

\subsubsection{ Minimum Pairwise Correlation}\label{sec:NPC}
The approaches considered so far compute the total correlation of a variate against all data variates or all remaining variates.
However, this might potentially be an overkill.
The final strategy that we consider takes into account only pairwise correlations into variate order. More specifically, it seeks a variate order where (a) the total pairwise correlations in the sequence is minimized and (b) pairs of variates that have smaller correlations are considered first.
The intuition is that considering conflicting {\em pairs of variates} early may dominate the comparison process, rendering subsequent variates unnecessary. 
Let 
$ m = |P|$.
In order to prioritize the variate order with lower correlations early on, we 
compute the pairwise score of a sequence, $\pi \in \Pi_P$, of variates as follows ($\Pi_P$ denotes the set of all permutations):
\[
pairwise\_score(\pi) = \sum_{i=1}^{m - 1} corr(\pi^i, \pi^{i + 1}) \times (m - i).
\]
Given this, the selected order, ${\pi_k}$, is computed as follows:
\[
{\pi_k} = \mathop{ARGMIN}_{\pi \in \Pi_P} pairwise\_score(\pi).
\]

\subsection{Number of Tuple-Pair Comparisons}\label{sec:lvos}
So far we have discussed the impact of the variate orders on the number of {\em v-check}s, under the assumption that the number of tuple comparisons is fixed. In this section, we discuss whether variate ordering strategies can have impact also on the number of tuple-pair dominance checks, or {\em t-check}s.

As we discussed in Section~\ref{sec:related-works}, some skyline algorithms, such as SFS \cite{Chomicki03skylinewith} and SaLSa \cite{Salsa},  require pre-sorting of the tuples 
%
to help prioritize tuples with higher filtering power~\cite{Chomicki03skylinewith}:

\begin{definition}[Filtering power of a Tuple]\label{def:filtering_power}
Given a tuple $t \in D$, its filtering power denotes the number of other tuples in $D$ that it dominates. Filtering power, $pow$, of a tuple $t$ is given by
$
pow(t) = |\{u \in D \mid t \succ u\}|.
$
\diaend \end{definition}

Here we argue that variate ordering strategies can potentially boost the effectiveness of the pre-sorting. 
Let $T_{order}$ be a tuple ordering strategy as introduced in Section~\ref{sec:problem}.
We define the filtering power of $T_{order}$ as follows:

\begin{definition}[Filtering power of $T_{order}$]\label{def:filtering_power_torder}
%
Given a tuple ordering strategy $T_{order}$ for data set $D$, 
%
\replaceieeecr{W}{w}e define the filtering power of the tuple ordering strategy $T_{order}$ as the
\[pow(T_{order}) = \sum_{t_i \in D} \left|\left\{t_j \in D\; s.t.\;\left(j>i\right)\wedge \left(t_i \succ t_j\right)\right\}\right|,
\]
where $i$ and $j$ are tuple ranks in $T_{order}$.


\diaend \end{definition}

In other words, generally speaking, tuples with high filtering power should come early (i.e., with low indexes) in the  tuple ordering strategy $T_{order}$ and prior works, including SFS~\cite{Chomicki03skylinewith} and SaLSa~\cite{Salsa}, have shown that any monotone order ensures that all tuples dominated by a given tuple come after that tuple in the tuple order ensuring maximum pruning power. This, however, does not mean that all monotone tuple orders have the same {\em pruning efficiency}, which also depends on other factors, such as the number of  tuples in the candidate set that need to be considered before the tuple can be excluded.

\subsubsection{Variate Orders and SFS/SaLSa}
\label{sec:variate_order_sfs_salsa}
Our key observation in this section is that the variate ordering strategy can  have a direct and significant impact on the filtering efficiency of the  tuple ordering strategy, $T_{order}$, if the ordering strategy itself is variate order sensitive as in lexicographical sort
\footnote{Lexicographic sorting is common in skylines~\cite{Salsa, Chomicki03skylinewith} and  other Pareto-front search algorithms~\cite{zhang2014efficient}.}
. Let us assume that the tuple order $T_{order}$ is monotone, which implies that once a tuple is seen and compared to the tuples that came before it, it cannot be pruned anymore. For such a tuple order, filtering efficiency is inversely proportional with the number of \replaceieeecr{"}{``}candidates\replaceieeecr{"}{''} each tuple is compared against before it can be pruned away (or confirmed as a skyline tuple):
\[
max\_cost_{t\text{-}check}(T_{order}) = \sum_{t_i \in D} |skyline\_tuples(i)|,
\]
where $|skyline\_tuples(i)|$ is the number of skyline elements produced {\bf before} seeing the $i^{th}$ tuple in $T_{order}$.
Importantly though the tuple $t_i$ may not need to be compared against all tuples  in $skyline\_tuples(i)$ to be refuted as a skyline tuple: scanning of the current list of skyline tuples can be stopped as soon as one tuple that dominates  $t_i$ is found among $skyline\_tuples(i)$.

As we experimentally validate in Section~\ref{sec:exp}, 
we argue\replaceieee{that}{}
\begin{enumerate}[leftmargin=*]
\item enumerating tuples in lexicographical order for the {\em min-correlation-first} variate order (according to the corresponding preference criteria), and 
\item comparing 
$t_i$
to the current list, $skyline\_tuples(i)$, of skyline tuples in a last-in-first-out (LIFO) manner 
\end{enumerate}
provides significant gains in filtering efficiency.
This is because (a) lexicographical sorting provides monotonicity, ensuring maximum filtering power for the tuple order, while (b) the last-in-first-out scanning of the current list of skyline objects sorted {\em in reverse order (i.e., LIFO)} in {\em min-correlation-first} variate order ensures that $t_i$ can be pruned with as few tuple checks ({\em t-check}s) as possible:

\begin{itemize}[leftmargin=*]
\item Reverse order scanning of the lexicographically produced skyline list ensures that the $t_i$ is compared to fewer skyline tuples as its dominators are often skyline tuples that were inserted recently, not necessarily the very earliest skyline tuples: this is because tuples closer in the lexicographical order are often more comparable to $t_i$ and thus more likely to dominate it than skyline tuples identified much earlier.

\item Because the first attribute is weakly correlated with the rest, in {\em min-correlation-first} variate order, lexicographical sort tends to expose skyline tradeoffs early: tuples that are dominating in the least-correlated attribute but vary substantially in the others are enumerated early and many of these are included in the skyline list because they are not easily dominated across all dimensions. 
\end{itemize}
Critically, {\em min-correlation-first} lexicographical sorting makes the LIFO order align well with the local dominance structure: 
with LIFO, $t_i$ is first compared against skyline tuples that are lexicographically close predecessors and those are often the best candidates to dominate $t_i$; if one of them dominates, then $t_i$ can be pruned after only a few {\em t-check}s.
%
%
In contrast,  {\em max-correlation-first} lexicographical sort, where the leading attribute is highly predictive of the others, may be better for finding dominant tuples earlier, but under LIFO, those tuples are checked later, not earlier, leading to a larger number of {\em t-checks}. Even if we revert the skyline comparison order from LIFO to FIFO, lacking the benefit of the local dominance structure, finding the dominating skyline tuple requires a larger number of {\em t-checks} increasing the overall cost. 

%

Given the above, 
we
introduce
\textit{Lex-Sorted Variate Ordering Skylines} (L-VOS),
in which tuples are lexicographically sorted according to {\em min-correlation-first} variate order. As we see in Section~\ref{sec:experiment_results}, this strategy performs better than the lexicographic order without considering the variate orders, as well as vanilla sort-first skylines (SFS)~\cite{Chomicki03skylinewith}, SaLSa~\cite{Salsa}.
%

\subsubsection{Variate Orders and D\&C/BBS}

As discussed in Section~\ref{sec:dps}, unlike SFS and \replacenewa{SALSA}{SaLSa}, D\&C~\cite{Borzsonyi01theskyline} and BBS~\cite{PapadiasTFS05} algorithms do not rely on pre-sorting of tuples; instead both rely on data partitioning and for both the order in which the tuples are considered and pruned indirectly depends on the variate order.

In particular, the divide-and-conquer (D\&C) skyline algorithm recursively splits the data by one attribute, computes local skylines, and then merges them while further recursively partitioning along other dimensions to reduce dominance checks and remove local skyline points dominated by skyline points from other partitions. In this process, the variate order affects performance by determining the order of splitting, thereby affecting the efficiency of the recursive subproblems. If we first split on a highly correlated attribute, later split dimensions may not add much additional discrimination; in contrast, with {\em min-correlation-first},  early split dimensions capture variation in data less explained by the other dimensions; consequently,  recursions along new attributes can contribute additional information, eliminating wasted work. 
%
%
BBS, however, does not directly benefit from variate ordering for {\em t-check} reduction: 
neither the shape of the R-tree, nor the pruning process depends in a predictable manner on the variate order.

Experiments in Section~\ref{sec:exp} confirm that the {\em min-correlation-first} strategy helps reduce the number of tuple-pair dominance checks ({\em t-check}s) for D\&C \replacenewa{}{\replaceieee{as well as}{and} BBS algorithms, \replaceieee{though}{but,} in terms of execution time, D\&C sees \replaceieee{hgher}{higher} gains.}  


\section{SIMD Vectorization}
\label{Sec:simd}

As discussed in Section~\ref{sec:correlations_and_dominance_checks}, the proposed variate ordering strategies
can reduce the average number of per-variate comparisons per dominance check.
While, as we experimentally evaluate in Section~\ref{sec:exp}, this can provide significant savings in conventional scalar architectures that process one operation on only one data element at a time, {\em single instruction, multiple data} (SIMD) architectures, which are able to process a single operation across multiple data elements simultaneously, need careful handling.

As shown in~\cite{10.1145/1893173.1893176}, skyline computations can benefit from SIMD architectures through the vectorization operation, where the dominance check is applied across multiple data dimensions simultaneously. In particular, assuming that all the preference attributes fit into a single SIMD vector (and organized contiguously in the memory for efficient vectorization), the dominance check operation can be done through a single SIMD operation, significantly reducing the skyline computation cost. In practice, however, SIMD hardware have limited bandwidth; for example a SIMD AVX-256, the industry standard for high-performance computing on x86 processors, can hold only 4 entries when the data type is 64 bit float. This implies that when the number of preference attributes is larger, the data has to be vectorized and processed in chunks and that early stopping through careful variate order selection can help reduce the number of per-chunk dominance checks (or {\em c-check}s). For instance, when the number of preference attributes is 5, in the above example with SIMD AVX-256 architecture, early stopping can reduce the skyline computation costs by $\sim 50\%$ simply by limiting the dominance check to one SIMD vector comparison, rather than two comparisons.  

We evaluate the computation time gains by VOS in SIMD architectures in Section~\ref{sec:exp}.




\section{Complexity Analysis}
There are four main steps for VOS  methods: 
\begin{enumerate}[left=0.3in]
    \item[\underline{\em Step 1}] Compute \replaceieee{the}{} correlation between preference attributes, \label{step:compute_corr}
    \item[\underline{\em Step 2}] Compute the best variate order,
    \item[\underline{\em Step 3}] Re-arrange the variates,
    \item[\underline{\em Step 3*}] Lexicographically sort the data,
    \label{step:lexsort}
    \item[\underline{\em Step 4}] Compute the skyline on the data\replaceieee{ using a skyline algorithm}{}. \label{step:compute_skyline}
\end{enumerate}

To perform \textit{Step 1}, we need the pairwise correlation of the attributes of the data in our dataset. This operation has a time complexity of $O(Nm^2)$, where $N$ is the data set size and $m$ is the number of preference attributes (or chunks in the case of SIMD hardware). 
Next, we identify the variate order according to the selected variate ordering strategy (Step 2). The time complexity of this operation depends on the strategy:
For  \textit{min-abs-correlation-first} (Section~\ref{sec:NACF}) and \textit{min-correlation-first} (Section~\ref{sec:NCF}), the time complexity to find the variate order is linear ($O(m)$) in the size of the preference set. 
%
For  \textit{inc-min-abs-correlation-first} (Section~\ref{sec:GNCF}) and \textit{inc-max-correlation-last} (Section~\ref{sec:GXCL}), the time complexity is quadratic ($O(m^2)$) in the size of the preference set. 
%
For \textit{min-pairwise-correlation} (Section \ref{sec:NPC}),  complexity increases exponentially, $O(m!)$, with the size of the preference set. 
%
%
Note that in practice we have $m \ll N$; moreover, in most cases we also have $m! \ll Nm^2$. Therefore, \replaceieeecr{}{for small values of $m$,}
the  $m!$  term in the {\em min-pairwise-correlation}  strategy can \replaceieeecr{generally}{}  be ignored.

\textit{Step 3} is not explicitly implemented; instead it is considered implicitly during the lexicographic sorting in Step 3* and the execution of the skyline algorithm in Step 4.
%
For L-VOS, there is an additional step (\textit{Step 3*}) where the data is pre-sorted before \textit{Step 4} is applied. This sorting operation has a time complexity of $O(NlogN)$, where $N$ is the number of tuples in the dataset. 
The time complexity of the final step, \textit{Step 4}, depends on the skyline algorithm being used (the skyline algorithms are polynomial in terms $m$ and $N$). 




\begin{table}[t]
\caption{\replaceieeecr{}{Experiment parameters (defaults are in bold)}
}\label{tab:experiment_parameters}
\vspace*{-0.05in}
\begin{adjustbox}{width=\columnwidth}
\begin{tabular}{|l|c|}\hline
{\bf Parameter}&{\bf Values}\\ \hline\hline
Number of preference attributes ($|P|$) & 3, \replaceieee{}{4,} \textbf{5}\replaceieee{, 7}{}\replaceieeecr{}{, 6, 7}\\ \hline
Number of Tuples & 50K, \textbf{100K}, 200K\\ \hline
Lexicographic Sorting (for SFS and SALSA)& {\bf True}, False\\ \hline

Rank of the $Cov_P$ Matrix & $1, \ldots, \boldsymbol{\lceil0.75|P|\rceil}, \ldots , |P|$ \\ \hline
Hardware & \textbf{Scalar}, SSE, AVX-256 \\ \hline
\end{tabular}
\end{adjustbox}
\vspace*{-0.05in}
\end{table}

\begin{table}[t]
\caption{Real-World Datasets \cite{Hindy_2021, uci_ml_repo}}
\label{tab:real_world_datasets}
\vspace*{-0.05in}
\begin{adjustbox}{width=\columnwidth}
\begin{tabular}{|l|c|c|}\hline
{\bf Name}&{\bf \# Variates}&{\bf Original \# of Tuples}\\ \hline
Concrete Compressive Strength & 9 & 1030\\ \hline
Hong Kong Weather & 10 & 18262\\ \hline
Individual Household Power Consumption & 9 & 2075259\\ \hline
Seoul Bike Demand  & 13 & 8760\\ \hline
Wine Quality (White) & 11 & 4898\\ \hline
\end{tabular}
\end{adjustbox}
\end{table}

\begin{table}[t]
  \centering
  \caption{Alternative variate ordering strategies and baselines considered in the experiments (note that in the acronyms we use \replaceieeecr{"}{``}N\replaceieeecr{"}{''} to denote \replaceieeecr{"}{``}minimum\replaceieeecr{"}{''} and \replaceieeecr{"}{``}X\replaceieeecr{"}{''} to denote \replaceieeecr{"}{``}maximum\replaceieeecr{"}{''})}
  \small
  \vspace*{-0.05in}

    \begin{tabular}{|l|l|}\hline
    \textbf{Best} & Empirically best variate order among all variate orders \\\hline
    \textbf{NACF} & Minimum Absolute Correlation First \\
    \textbf{NCF} & Minimum Correlation First \\
    \textbf{INCF} & Incremental Minimum Correlation First \\
    \textbf{IXCL} & Incremental Maximum Correlation Last \\
    \textbf{NPC} & Minimum Pairwise Correlation  \\\hline
    \textbf{Vanilla} & Average of all possible variate orders for vanilla skyline\\
    \textbf{XCF} & Maximum Correlation First \\ 
     \textbf{Worst} & Empirically worst variate order among all variate orders \\ \hline
    \end{tabular}%
  \label{tab:confs}%
  \vspace*{-0.05in}
\end{table}%

\begin{table}[t]
 \caption{\replaceieee{}{Exec. times of the baseline algorithms (\replacenewa{200K}{100K} data, \replaceieee{5 attributes}{$|P|=5$, $Cov_P$ matrix rank = 4})}}\label{tab:vanilla}
 \vspace*{-0.05in}
  \centering
  \begin{tabular}{|l|c|c|c|c|c|}
    \hline
    \multicolumn{6}{|c|}{\textbf{Avg. Exec. Time of  Vanilla Skyline Algorithms (seconds)}} \\
    \hline
    & \textbf{BNL} & \textbf{SFS} & \textbf{SaLSa} & \textbf{D\&C} & \textbf{BBS} \\
    \hline
    \textbf{Synthetic Data} & 6.97 & 3.06 & 3.16 & 0.16 & 9.20 \\
    \hline
    \textbf{Real Data} & 0.15 & 0.10 & 0.19 & 0.09 & 0.28 \\
    \hline
  \end{tabular}
\end{table}

\begin{table}[t]
  \caption{\replaceieee{}{Exec. times of variate ordering strategies (listed in Table~\ref{tab:confs}) (100K data, $|P|=5$, $Cov_P$ matrix rank $=4$)}}\label{tab:vo_strategies_exec_time}
  \vspace*{-0.05in}
  \centering
  \adjustbox{width=\columnwidth}{
    \begin{tabular}{|l|c|c|c|c|c|c|}
        \hline
        \multicolumn{7}{|c|}{\textbf{Average Execution Time for Variate Ordering Strategies (seconds)}} \\
        \hline
        \textbf{VO Strategy} & \textbf{NCF} & \textbf{NACF} & \textbf{INCF} & \textbf{IXCL} & \textbf{NPC} & \textbf{XCF} \\
        \hline
        \textbf{Synthetic Data} & 0.0066 & 0.0066 & 0.0067 & 0.0067 & 0.0068 & 0.0066 \\
        \hline
        \textbf{Real Data} & 0.0065 & 0.0065 & 0.0066 & 0.0066 & 0.0067 & 0.0065 \\
        \hline
    \end{tabular}
    \vspace*{-0.1in}
  }
\end{table}


\begin{table}[t]
    \caption{\replaceieee{}{Exec. time gains provided by the alternative variate ordering strategies relative to vanilla skyline algorithms  (100K, synthetic data, $|P|$ = 5; the table is colored per-column, from best/green to worst/red)}}
  \label{tab:vos_alternatives}%
  \vspace*{-0.05in}
  \centering
    \begin{tabular}{|c|ccccc|}
    \hline
          & \multicolumn{5}{c|}{\cellcolor[rgb]{ .851,  .851,  .851}\textbf{Execution Time Gains - (100K Synthetic Data)}} \\
          & \cellcolor[rgb]{ .851,  .851,  .851}\textbf{BNL-NL} & \cellcolor[rgb]{ .851,  .851,  .851}\textbf{SFS-L} & \cellcolor[rgb]{ .851,  .851,  .851}\textbf{SaLSa-L} & \cellcolor[rgb]{ .851,  .851,  .851}\textbf{D\&C-NL} & \cellcolor[rgb]{ .851,  .851,  .851}\textbf{BBS-NL} \\
    \hline
    \rowcolor[rgb]{ .384,  .749,  .482} \textit{\textbf{Best}} & \cellcolor[rgb]{ .388,  .745,  .482}\textit{20\%} & \cellcolor[rgb]{ .388,  .745,  .482}\textit{32\%} & \cellcolor[rgb]{ .388,  .745,  .482}\textit{47\%} & \cellcolor[rgb]{ .388,  .745,  .482}\textit{54\%} & \cellcolor[rgb]{ .388,  .745,  .482}\textit{14\%} \\
    \hline
    \multicolumn{1}{|r|}{\textbf{NCF}} & \cellcolor[rgb]{ .871,  .941,  .898}6\% & \cellcolor[rgb]{ .976,  .984,  .988}24\% & \cellcolor[rgb]{ .984,  .988,  .996}41\% & \cellcolor[rgb]{ .522,  .8,  .6}47\% & \cellcolor[rgb]{ .784,  .906,  .827}8\% \\
        \multicolumn{1}{|r|}{\textbf{IXCL}} & \cellcolor[rgb]{ .667,  .859,  .722}12\% & \cellcolor[rgb]{ .729,  .886,  .776}27\% & \cellcolor[rgb]{ .773,  .902,  .812}43\% & \cellcolor[rgb]{ .557,  .816,  .627}46\% & \cellcolor[rgb]{ .918,  .961,  .941}6\% \\
    \multicolumn{1}{|r|}{\textbf{NACF}} & \cellcolor[rgb]{ .984,  .918,  .929}-3\% & \cellcolor[rgb]{ .984,  .98,  .992}23\% & \cellcolor[rgb]{ .984,  .984,  .996}41\% & \cellcolor[rgb]{ .984,  .855,  .867}-15\% & \cellcolor[rgb]{ .98,  .831,  .839}-3\% \\
    \multicolumn{1}{|r|}{\textbf{INCF}} & \cellcolor[rgb]{ .984,  .945,  .957}-1\% & \cellcolor[rgb]{ .984,  .949,  .961}21\% & \cellcolor[rgb]{ .984,  .949,  .961}40\% & \cellcolor[rgb]{ .569,  .82,  .639}45\% & \cellcolor[rgb]{ .761,  .898,  .804}8\% \\
    \multicolumn{1}{|r|}{\textbf{NPC}} & \cellcolor[rgb]{ .714,  .878,  .765}11\% & \cellcolor[rgb]{ .733,  .886,  .78}27\% & \cellcolor[rgb]{ .761,  .898,  .804}43\% & \cellcolor[rgb]{ .984,  .918,  .929}4\% & \cellcolor[rgb]{ .984,  .965,  .976}4\% \\
    \multicolumn{1}{|r|}{\textbf{XCF}} & \cellcolor[rgb]{ .98,  .816,  .827}-11\% & \cellcolor[rgb]{ .984,  .98,  .992}23\% & \cellcolor[rgb]{ .984,  .973,  .984}40\% & \cellcolor[rgb]{ .976,  .584,  .592}-95\% & \cellcolor[rgb]{ .976,  .686,  .694}-10\% \\
    \hline
    \rowcolor[rgb]{ .973,  .408,  .42} \textit{\textbf{Worst}} & \cellcolor[rgb]{ .973,  .412,  .42}\textit{-45\%} & \cellcolor[rgb]{ .973,  .412,  .42}\textit{-6\%} & \cellcolor[rgb]{ .973,  .412,  .42}\textit{27\%} & \cellcolor[rgb]{ .973,  .412,  .42}\textit{-147\%} & \cellcolor[rgb]{ .973,  .412,  .42}\textit{-24\%} \\
    \hline
    \end{tabular}%
\end{table}%


\begin{table}[t]
    \caption{\replaceieee{}{Exec. time gains provided by the alternative variate ordering strategies relative to vanilla skyline algorithms  (100K, real data, 5 attributes;  the table is colored per-column, from best/green to worst/red)}}
  \vspace*{-0.05in}
  \centering
    \begin{tabular}{|c|ccccc|}
    \hline
          & \multicolumn{5}{c|}{\cellcolor[rgb]{ .851,  .851,  .851}\textbf{Execution Time Gains - (100K Real Data)}} \\
          & \cellcolor[rgb]{ .851,  .851,  .851}\textbf{BNL-NL} & \cellcolor[rgb]{ .851,  .851,  .851}\textbf{SFS-L} & \cellcolor[rgb]{ .851,  .851,  .851}\textbf{SaLSa-L} & \cellcolor[rgb]{ .851,  .851,  .851}\textbf{D\&C-NL} & \cellcolor[rgb]{ .851,  .851,  .851}\textbf{BBS-NL} \\
    \hline
    \rowcolor[rgb]{ .384,  .749,  .482} \textit{\textbf{Best}} & \cellcolor[rgb]{ .388,  .745,  .482}\textit{12\%} & \cellcolor[rgb]{ .388,  .745,  .482}\textit{29\%} & \cellcolor[rgb]{ .388,  .745,  .482}\textit{47\%} & \cellcolor[rgb]{ .388,  .745,  .482}\textit{39\%} & \cellcolor[rgb]{ .388,  .745,  .482}\textit{10\%} \\
    \hline
    \multicolumn{1}{|r|}{\textbf{NCF}} & \cellcolor[rgb]{ .851,  .933,  .882}5\% & \cellcolor[rgb]{ .949,  .973,  .965}21\% & \cellcolor[rgb]{ .973,  .984,  .988}41\% & \cellcolor[rgb]{ .71,  .875,  .761}26\% & \cellcolor[rgb]{ .773,  .902,  .816}7\% \\
    \multicolumn{1}{|r|}{\textbf{IXCL}} & \cellcolor[rgb]{ .698,  .871,  .749}7\% & \cellcolor[rgb]{ .929,  .965,  .949}21\% & \cellcolor[rgb]{ .957,  .976,  .973}41\% & \cellcolor[rgb]{ .859,  .937,  .89}20\% & \cellcolor[rgb]{ .745,  .89,  .788}8\% \\
    \multicolumn{1}{|r|}{\textbf{NACF}} & \cellcolor[rgb]{ .984,  .922,  .933}0\% & \cellcolor[rgb]{ .984,  .953,  .965}19\% & \cellcolor[rgb]{ .984,  .953,  .965}40\% & \cellcolor[rgb]{ .984,  .941,  .953}9\% & \cellcolor[rgb]{ .984,  .863,  .875}0\% \\
    \multicolumn{1}{|r|}{\textbf{INCF}} & \cellcolor[rgb]{ .984,  .918,  .925}0\% & \cellcolor[rgb]{ .984,  .961,  .973}19\% & \cellcolor[rgb]{ .984,  .961,  .973}40\% & \cellcolor[rgb]{ .741,  .89,  .788}24\% & \cellcolor[rgb]{ .953,  .976,  .973}6\% \\
    \multicolumn{1}{|r|}{\textbf{NPC}} & \cellcolor[rgb]{ .71,  .875,  .761}7\% & \cellcolor[rgb]{ .922,  .965,  .945}21\% & \cellcolor[rgb]{ .969,  .98,  .98}41\% & \cellcolor[rgb]{ .98,  .808,  .82}-7\% & \cellcolor[rgb]{ .984,  .98,  .992}6\% \\
    \multicolumn{1}{|r|}{\textbf{XCF}} & \cellcolor[rgb]{ .98,  .729,  .737}-6\% & \cellcolor[rgb]{ .984,  .969,  .98}20\% & \cellcolor[rgb]{ .984,  .976,  .988}41\% & \cellcolor[rgb]{ .973,  .553,  .561}-37\% & \cellcolor[rgb]{ .976,  .682,  .69}-8\% \\
    \hline
    \rowcolor[rgb]{ .973,  .408,  .42} \textit{\textbf{Worst}} & \cellcolor[rgb]{ .973,  .412,  .42}\textit{-17\%} & \cellcolor[rgb]{ .973,  .412,  .42}\textit{0\%} & \cellcolor[rgb]{ .973,  .412,  .42}\textit{28\%} & \cellcolor[rgb]{ .973,  .412,  .42}\textit{-54\%} & \cellcolor[rgb]{ .973,  .412,  .42}\textit{-21\%} \\
    \hline
    \end{tabular}%
  \label{tab:vos_alternatives_real}%
  \vspace*{-0.05in}
\end{table}%

\begin{table}[t]
\caption{\replaceieee{}{Reductions in {\em t-check}, {\em v-check}, and execution time relative to vanilla skylines for varying \replaceieeecr{strategies}{number of tuples} \replaceieeecr{}{(synthetic data, $|P| = 5$) \underline{(higher is better)}}}}
\vspace*{-0.05in}
\centering{%
\adjustbox{width=\columnwidth}{
\begin{tabular}{c}
    \begin{tabular}{|r|ccccc|}
    \hline
          & \multicolumn{5}{c|}{\cellcolor[rgb]{ .851,  .851,  .851}\textbf{\% reduction in t-checks for NCF wrt. Vanilla-NL}}\\
          & \cellcolor[rgb]{ .851,  .851,  .851}\textbf{BNL-NL} & \cellcolor[rgb]{ .851,  .851,  .851}\textbf{SFS-L} & \cellcolor[rgb]{ .851,  .851,  .851}\textbf{SaLSa-L} & \cellcolor[rgb]{ .851,  .851,  .851}\textbf{D\&C-NL} & \cellcolor[rgb]{ .851,  .851,  .851}\textbf{BBS-NL} \\
    \hline
    \textbf{50K} & 0\%   & 51\%  & 56\%  & 46\%  & -1\% \\
    \textbf{100K} & 0\%   & 55\%  & 58\%  & 48\%  & -4\% \\
    \textbf{200K} & 0\%   & 58\%  & 55\%  & 50\%  & -3\%\\
    \hline
    \end{tabular}\\
\replaceieee{}{(a) reductions in {\em t-checks}}\\ \\
    \begin{tabular}{|r|ccccc|}
    \hline
          & \multicolumn{5}{c|}{\cellcolor[rgb]{ .851,  .851,  .851}\textbf{\% reduction in v-checks for NCF wrt. Vanilla-NL}}\\
          & \cellcolor[rgb]{ .851,  .851,  .851}\textbf{BNL-NL} & \cellcolor[rgb]{ .851,  .851,  .851}\textbf{SFS-L} & \cellcolor[rgb]{ .851,  .851,  .851}\textbf{SaLSa-L} & \cellcolor[rgb]{ .851,  .851,  .851}\textbf{D\&C-NL} & \cellcolor[rgb]{ .851,  .851,  .851}\textbf{BBS-NL} \\
    \hline
    \textbf{50K} & 8\%   & 45\%  & 33\%  & 41\%  & 41\% \\
    \textbf{100K} & 8\%   & 49\%  & 35\%  & 42\%  & 39\% \\
    \textbf{200K} & 8\%   & 52\%  & 29\%  & 44\%  & 41\%\\
    \hline
    \end{tabular}\\
\replaceieee{}{(b) reductions in {\em v-checks}}\\ \\
    \begin{tabular}{|r|ccccc|}
    \hline
          & \multicolumn{5}{c|}{\cellcolor[rgb]{ .851,  .851,  .851}\textbf{\% reduct. in exec. time for NCF wrt. Vanilla-NL}}\\
          & \cellcolor[rgb]{ .851,  .851,  .851}\textbf{BNL-NL} & \cellcolor[rgb]{ .851,  .851,  .851}\textbf{SFS-L} & \cellcolor[rgb]{ .851,  .851,  .851}\textbf{SaLSa-L} & \cellcolor[rgb]{ .851,  .851,  .851}\textbf{D\&C-NL} & \cellcolor[rgb]{ .851,  .851,  .851}\textbf{BBS-NL}\\
    \hline
    \textbf{50K} & 7\%   & 20\%  & 39\%  & 46\%  & 10\% \\
    \textbf{100K} & 6\%   & 24\%  & 41\%  & 47\%  & 8\% \\
    \textbf{200K} & 7\%   & 24\%  & 40\%  & 50\%  & 8\% \\
    \hline
    \end{tabular}\\
\replaceieee{}{(c) reductions in execution times}\\
\end{tabular}
}}
\label{tab:varsize}
\end{table}

\begin{table}[t]
    \caption{
    \replaceieeecr{}{Impact of the number of preference variates on skyline computation time}
    \underline{(higher is better)} -- 100K, default 
    }
    \vspace*{-0.05in}
    \centering
    \begin{tabular}{|r|ccccc|}
    \hline
          & \multicolumn{5}{c|}{\cellcolor[rgb]{ .851,  .851,  .851}\textbf{\% reduct. in exec. time for NCF wrt. Vanilla-NL}} \\
          & \cellcolor[rgb]{ .851,  .851,  .851}\textbf{BNL-NL} & \cellcolor[rgb]{ .851,  .851,  .851}\textbf{SFS-L} & \cellcolor[rgb]{ .851,  .851,  .851}\textbf{SaLSa-L} & \cellcolor[rgb]{ .851,  .851,  .851}\textbf{D\&C-NL} & \cellcolor[rgb]{ .851,  .851,  .851}\textbf{BBS-NL} \\
    \hline
    \textbf{$\mathbf{|P| = 3}$} & 3\%   & 11\%  & 19\%  & 18\%  & 1\% \\
    \textbf{$\mathbf{|P| = 4}$} & 7\%   & 17\%  & 28\%  & 46\%  & 5\% \\ 
    \textbf{$\mathbf{|P| = 5}$} & 6\%   & 24\%  & 41\%  & 47\%  & 8\% \\
    \textbf{$\mathbf{|P| = 6}$} & 9\%   & 23\%  & 52\%  & 51\%  & 11\% \\
    \textbf{$\mathbf{|P| = 7}$} & 8\%   & 12\%  & 54\%  & 48\% & 15\% \\
    \hline
    \end{tabular}%
    \label{tab:ablation_1}
    \vspace*{-0.05in}
\end{table}

\begin{table}[t]
    \caption{
    \replaceieeecr{}{
    Impact of the number of preference variates on skyline computation time, with  conditioned standard deviation of correlations ($0.2 < STD < 0.3$) 
    \underline{(higher is better)} -- 100K, default 
    }}
    \vspace*{-0.05in}
    \centering
    \begin{tabular}{|r|ccccc|}
    \hline
          & \multicolumn{5}{c|}{\cellcolor[rgb]{ .851,  .851,  .851}\textbf{\% reduct. in exec. time for NCF wrt. Vanilla-NL}} \\
          & \cellcolor[rgb]{ .851,  .851,  .851}\textbf{BNL-NL} & \cellcolor[rgb]{ .851,  .851,  .851}\textbf{SFS-L} & \cellcolor[rgb]{ .851,  .851,  .851}\textbf{SaLSa-L} & \cellcolor[rgb]{ .851,  .851,  .851}\textbf{D\&C-NL} & \cellcolor[rgb]{ .851,  .851,  .851}\textbf{BBS-NL} \\
    \hline
    \textbf{$\mathbf{|P| = 3}$} & 3\% & 2\% & 8\% &  20\% & 1\% \\
    \textbf{$\mathbf{|P| = 4}$} &  6\%  &  24\% & 38\%  &  44\% & 5\% \\
    \textbf{$\mathbf{|P| = 5}$} &  7\%  &  24\% &  39\% & 51\%  & 8\% \\
    \textbf{$\mathbf{|P| = 6}$} &  13\%  &  48\% & 63\%  & 61\%  & 17\% \\
    \textbf{$\mathbf{|P| = 7}$} &  14\%  &  54\% &  70\% & 63\% & 21\% \\
    \hline
    \end{tabular}
    \label{tab:ablation_1_conditioned}
\end{table}


\begin{table}[t]
    \caption{\replaceieee{}{Impact of the rank of the correlation matrix on skyline computation time \underline{(higher is better)} -- 100K, default}}
    \vspace*{-0.05in}
  \centering
    \begin{tabular}{|r|ccccc|}
    \hline
          & \multicolumn{5}{c|}{\cellcolor[rgb]{ .851,  .851,  .851}\textbf{\% reduct. in exec. time for NCF wrt. Vanilla-NL}}\\
          & \cellcolor[rgb]{ .851,  .851,  .851}\textbf{BNL-NL} & \cellcolor[rgb]{ .851,  .851,  .851}\textbf{SFS-L} & \cellcolor[rgb]{ .851,  .851,  .851}\textbf{SaLSa-L} & \cellcolor[rgb]{ .851,  .851,  .851}\textbf{D\&C-NL} & \cellcolor[rgb]{ .851,  .851,  .851}\textbf{BBS-NL} \\
    \hline
    \textbf{Rank 1} & 0\%   & 63\%  & 79\%  & 16\%  & 8\% \\
    \textbf{Rank 2} & 6\%   & 16\%  & 35\%  & 45\%  & 10\% \\
    \textbf{Rank 3} & 8\%   & 20\%  & 35\%  & 53\%  & 10\% \\
    \textbf{Rank 4} & 6\%   & 24\%  & 41\%  & 47\%  & 8\% \\
    \textbf{Rank 5} & 8\%   & 13\%  & 35\%  & 43\%  & 6\% \\
    \hline
    \end{tabular}%
    \vspace*{-0.05in}
  \label{tab:rank}%
\end{table}%


\begin{table}[t]
\caption{\replaceieee{}{Performance improvements ($\%$ reductions) in \replacenewa{L-}{}VOS for different hardware configurations}}
\label{tab:simd_results}
\vspace*{-0.05in}
\centering{%
   \begin{tabular}{c}
    \begin{tabular}{|r|ccccc|}
    \hline
          & \multicolumn{5}{P{25em}|}{\cellcolor[rgb]{ .851,  .851,  .851}\textbf{\% reduction in v- or c-checks for NCF wrt. Vanilla-NL}}\\
          & \cellcolor[rgb]{ .851,  .851,  .851}\textbf{BNL-NL} & \cellcolor[rgb]{ .851,  .851,  .851}\textbf{SFS-L} & \cellcolor[rgb]{ .851,  .851,  .851}\textbf{SaLSa-L} & \cellcolor[rgb]{ .851,  .851,  .851}\textbf{D\&C-NL} & \cellcolor[rgb]{ .851,  .851,  .851}\textbf{BBS-NL} \\
    \hline
    \textbf{Scalar} & 8\%   & 49\%  & 35\%  & 42\%  & 39\% \\
    \textbf{SSE} & 6\%   & 53\%  & 40\%  & 37\%  & 24\% \\
    \textbf{AVX} & 3\%   & 55\%  & 46\%  & 42\%  & 3\%\\
    \hline
    \end{tabular}%
    \\    \replaceieee{}{(a) v-check or c-check}\\\\
    \begin{tabular}{|r|ccccc|}
    \hline
          & \multicolumn{5}{P{25em}|}{\cellcolor[rgb]{ .851,  .851,  .851}\textbf{\% reduct. in exec. time for NCF wrt. Vanilla-NL}} \\
          & \cellcolor[rgb]{ .851,  .851,  .851}\textbf{BNL-NL} & \cellcolor[rgb]{ .851,  .851,  .851}\textbf{SFS-L} & \cellcolor[rgb]{ .851,  .851,  .851}\textbf{SaLSa-L} & \cellcolor[rgb]{ .851,  .851,  .851}\textbf{D\&C-NL} & \cellcolor[rgb]{ .851,  .851,  .851}\textbf{BBS-NL} \\
    \hline
    \textbf{Scalar} & 6\%   & 24\%  & 41\%  & 47\%  & 8\% \\
    \textbf{SSE} & 4\%   & 20\%  & 37\%  & 47\%  & -2\% \\
    \textbf{AVX} & 5\%   & 22\%  & 35\%  & 48\%  & -3\% \\
    \hline
    \end{tabular}
    \\
    \replaceieee{}{(b) Execution time}
   \end{tabular}
}
\vspace*{0.05in}
\end{table}

\section{Experiments and Analysis}
In this section we present experiments that analyze the performance of variate ordering strategies for skylines. We analyze how our proposed methods perform on varying dataset sizes and varying dimensionality of the preference space. 
The experiment parameters are presented in Table \ref{tab:experiment_parameters}.

%

Experiments are run on compute nodes \replaceieeecr{}{provided by Chameleon Cloud~\cite{keahey2020lessons}} with the following hardware configuration: AMD EPYC 7763 64-Core Processor \replaceieeecr{, with 2 sockets, 64 cores per socket, and 2 threads per core. The maximum CPU frequency is}{@} 3.529 GHz. The instance had 251.28 GiB of usable memory and 500 GB of SSD storage. We used Ubuntu 24.04 operating system and Python 3.12.7 to run our experiments.
%
All the skyline computation code is written in C and called through \replaceieeecr{python}{Python} using the ctypes library from PyPI \footnote{\replaceieeecr{}{Code available at - https://github.com/Emitlab/Variate-Ordering-Strategies-for-Skyline-Query-Optimization}}. The dominance check counts and runtime is computed within the C skyline computation shell. \replacenewa{}{We consider three hardware implementations of dominance checks:  we use a scalar implementation as the default and  experiment with vectorized dominance checks using SSE and AVX-256 hardware implementations of SIMD.}

%


\subsection{Experiment Setup}
\subsubsection{Datasets}
In these experiments, we use both real-world and synthetic datasets:
\begin{itemize}[leftmargin=*]
    \item \textbf{Real-World Datasets} - The real-world datasets we use are listed in Table \ref{tab:real_world_datasets}.
    \replaceieeecr{}{To reach the target data size, smaller datasets are oversampled via Gaussian-noise perturbation \cite{beinecke2021gaussian}; larger datasets are randomly under-sampled.}

    \item \textbf{Synthetic Datasets} - 
Table~\ref{tab:experiment_parameters} lists data set configurations used in these experiments. Importantly,
%
%
synthetic datasets are generated to cover the  spectrum of correlation profiles with varying ranks for each preference set size  \replaceieee{}{(a correlation matrix with low rank implies more structure within the data, whereas higher rank implies lesser structure).}
\replaceieee{}{To do this, we sample correlation matrices using different random seeds, and then sample data that fits this correlation matrix from a zero-mean unit-variance multi-variate normal distribution\footnote{We use  \texttt{numpy.random.multivariate\_normal()}.}.} 
\replaceieeecr{\replaceieee{Our dataset generation process is detailed}{More details of our dataset generation process, including the code, are given} in our \replaceieee{anonymous}{} github repository\footnote{
https://github.com/Emitlab/Variate-Ordering-Strategies-for-Skyline-Query-Optimization}.}{}


\end{itemize}  

%







\noindent Without loss of generality, we assume that the preference criteria for all attributes is maximization.
\replacenewa{}{To select the preference sets \replaceieee{}{for real-world data }per Table~\ref{tab:experiment_parameters}, we sample all possible three, five and seven variate preference sets and randomly select 10\% of the total number of preference sets for each $(|P|, rank)$ combination.} 
For both synthetic and real-world data sets, the data is generated (or augmented) multiple (5) times with different random seeds and we report averages.
\subsubsection{Skyline Algorithms}


 \noindent
 {\bf Block Nested Loop (BNL) Skyline \cite{Borzsonyi01theskyline}} algorithm maintains a window of skyline candidates. It   performs a pass over the data and for each tuple it checks dominance condition with all the tuples that are currently skyline candidates (and in the window). If any of the  candidates violates the dominance condition, it is removed from the window; if the new tuple is non-dominated by tuples currently in the window, it is added to the current skyline candidates. 

 \noindent
{\bf Sort Filter Skyline (SFS) \cite{Chomicki03skylinewith}} uses a pre-sorting based approach to reduce the number of dominance checks in the process of skyline search. SFS uses a monotonic function $\langle sum\rangle$ to perform this pre-sorting operation. 
%

 \noindent
{\bf Sort and Limit Skyline (SaLSa) \cite{Salsa}} 
algorithm extends upon SFS by interleaving partial sorting (using $\langle max, sum \rangle$) with dominance checks. This ensures that a tuple  removed from the skyline candidates set \replaceieee{can never}{cannot} be introduced into the set again.

     \noindent
{\bf Branch and Bound Skyline (BBS)~\cite{PapadiasTFS05}} 
algorithm indexes input data onto an R-Tree index using STR bulk loading~\cite{582015}. This enables the BBS algorithm to prune large subsets of data using minimum bounding boxes. 

     \noindent
{\bf Divide and Conquer (D\&C) Skyline~\cite{Borzsonyi01theskyline,KungLP75}} 
algorithm partitions the data incrementally, one variate at a time, along the variate's median and then computes the skylines for each of the partitions.  The algorithm merges these partitions in a bottom-up manner until we are left with one global skyline.

\replaceieee{Figure~\ref{fig:vanilla}}{Table~\ref{tab:vanilla}} shows the execution times of the baseline algorithms on 5 attribute, \replacenewa{200K}{100K} data set. Note that the BBS algorithm time \replaceieee{does not include the time to construct the underlying R-tree}{does not include the R-Tree construction time}.
\replaceieee{}{Interestingly, the table  \replaceieee{Figure~\ref{fig:vanilla}}{shows that BBS has the highest average execution time. This is due to outlier cases which lead to the entire dataset being in the skyline. In such cases, BBS is unable to take advantage of MBR-based pruning~\cite{PapadiasTFS05}.}
}

\subsubsection{Variants}
Table~\ref{tab:confs} lists the variate ordering variants considered.
As visualized at the top and bottom of this table, as additional yardsticks, we also consider the empirically best variate order, empirically worst variate order, and a \textit{maximum correlation first} variate ordering strategy, which based on the earlier arguments in the paper, we expect to perform worse than vanilla skyline.
\replaceieee{}{Table~\ref{tab:vo_strategies_exec_time} shows the execution time to compute optimal variate order using the different strategies listed in Table~\ref{tab:confs}. The results show that optimal variate order computation is significantly cheaper than skyline query execution time (Table~\ref{tab:vo_strategies_exec_time} vs. Table~\ref{tab:vanilla}).}
The default for SFS and SaLSa is the LIFO based lex-sort (SFS-L, SALSA-L), while \replacenewa{}{for} BNL, BBS, and D\&C algorithms, lex-sort has not been applied (BNL-NL, BBS-NL, D\&C-NL).
Since vanilla skylines are not variate order-aware, we ran all permutations of preference variates and report the average performance.

\subsubsection{Evaluation Criteria}
\phantom{.}
\replacenewa{}{
To evaluate the performance of variate ordering in skyline algorithms, we consider \textit{the total number of tuple-pair dominance checks {\em \replaceieee{}{(}t-checks)}}, 
\textit{the total number of per-variate dominance checks {\em \replaceieee{}{(}v-checks)}}, and \textit{the skyline computation time}.
We report average gains: for each scenario we compute  the $\%$ reduction in cost when using the proposed strategies with respect to the vanilla algorithms and we report the average of these $\%$ reductions.
}
%
%
%
Each experiment configuration is run \replaceieee{5}{100} times (with \replaceieee{}{5} random seeds where applicable) and we report averages.

\subsection{Experiment Results}\label{sec:experiment_results}\label{sec:exp}

In this section, we first provide an overview of how variate ordering strategies perform in terms of number of variate comparisons and the skyline computation time. 

\subsubsection{Overview Results}

\replaceieee{Figure~\ref{fig:vos_alternatives}}{Table~\ref{tab:vos_alternatives}} provides an overview of the execution time gains provided by the alternative variate ordering strategies relative to the vanilla skyline algorithms on synthetic data. The table also shows the execution times provided by the variate-order that provides the lowest execution time and the highest execution time  along with the {\em max-correlation-first} (XCF) baseline.
\begin{itemize}[leftmargin=*]
\item As we see in this table, \replacenewa{except for BBS}{while \replaceieee{all}{most} minimum correlation favoring strategies provide positive gains}, the 
{\em min-correlation-first} (NCF) and {\em incremental maximum correlation last} (IXCL) strategies provide the most consistent \replaceieee{highest}{} time gains, often close to the best gains possible.
\item \replaceieee{}{Minimum Absolute Correlation First (NACF) strategy provides negative gains for BNL, D\&C, and BBS skyline algorithms. This shows that pushing un-correlated variates first hurts the skyline computation efficiency. SaLSa and SFS get an overall benefit from NACF, since they use NCF-based lexicographical sorting as a pre-processing step.}
\item \replacenewa{}{Experiments also confirm that the maximum correlation favoring competitor, {\em max-correlation-first} (XCF) indeed performs poorly among all strategies and, except for the SFS and SaLSa algorithms that also benefit from NCF-based lexicographical sorting, results in negative gains, close to the worst case scenario.}
\end{itemize}
%
\replaceieee{}{Table~\ref{tab:vos_alternatives_real} with real data sets confirms these results.
For the rest of our experiments we use the {\em min-correlation-first} (NCF) strategy for all skyline algorithms as the results above show it is among the best performing strategies.
}
\replaceieee{Figure~\ref{fig:varsize}}{Table~\ref{tab:varsize}} further analyzes these results by diving deeper into {\em t-check}, {\em v-check}, and execution time gain of the NCF strategy for varying number of tuples and different skyline algorithms:
\begin{itemize}[leftmargin=*]
\item 
Performance gain results are consistent across data sizes.
\item Variate ordering does not provide any {\em t-check} gains for BNL and BBS, while SFS, SaLSa, and D\&C see significant reductions in number of tuple-pair dominant checks. \replacenewa{}{This is because, the output of the partioning step in D\&C and the pre-sorting done in SFS and SaLSa is dependent on the order of variates in the dataset and the strategies proposed in this paper provide significant pruning opportunities.}
\item All algorithms, except BNL, show significant\replaceieeecr{numbers of}{} reductions in the number of\replaceieee{per-attribute dominance checks}{} {\em v-checks}.
\item This drop in the number of {\em v-checks} results in execution time gains for all algorithms\replacenewa{, except for BBS}{. Among all algorithms, time gains for BNL and D\&C follow closely the reductions in {\em v-checks}; for the other three algorithms, time gains are lower due to other overheads.}
\end{itemize}

\begin{figure}[t]
 \centering{%
    \begin{tabular}{c}
    \includegraphics[width=0.55\columnwidth]{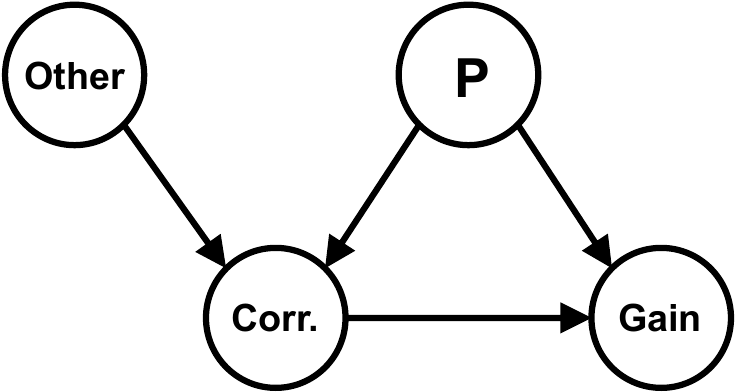}\\
    (a) Causal model of gains\\ \\
     \includegraphics[width=0.55\columnwidth]{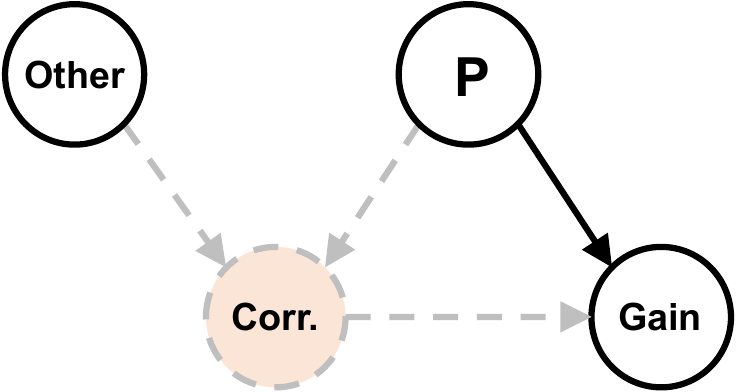}\\
     (b) Conditioned causal model of gains
    \end{tabular}
 }
 \caption{
 \replaceieeecr{}{(a) Number of preference attributes impacts gain both directly and through its impact on the data correlation; (b) conditioning on correlation reveals true impact of the number of preference attributes on gain} }
 \label{fig:causal}
  \vspace*{-0.05in}
 \end{figure}

\subsubsection{Impact of the Number of Preference Attributes}

As the number of preference attributes in the skyline problem increases, the problem tends to become costlier. This is because the search volume increases exponentially and, hence, the area of the Pareto-front also increases. 
In terms of using variate ordering to optimize skyline search, the number of possible variate orders also increases as the number of preference attributes for the data  increases. This in turn means that the strategies for ordering variates  have to be more accurate\replaceieee{ in their predictions}{}.



To study the impact of the number of preference attributes, we ran a set of experiments, changing the number of preference variates, while keeping all other default parameters listed in Table \ref{tab:experiment_parameters} constant. Results are presented in \replaceieee{}{Table~\ref{tab:ablation_1}}.
%
In the table, we observe that \replaceieeecr{}{for the synthetic data distribution we considered in this paper,} the time gains 
\replaceieeecr{}{initially increase and then generally stabilize. 
This behavior arises because of the way the  number of preference attributes causally impacts the gains. As we see in Figure~\ref{fig:causal}(a), the number of preference attributes impacts gain both directly and through its impact on the underlying data correlation. More specifically, we construct correlation matrices using vectors sampled uniformly from a unit sphere, with each pairwise correlation given by the dot product of the corresponding vectors. 
In higher dimensions, randomly sampled vectors tend to be nearly orthogonal, leading to correlations closer to zero and lower standard deviation among per-variate average correlations. 
Figure~\ref{fig:stdev_trend_varying_p} confirms this.

}

\replaceieeecr{}{


\begin{figure}[t]
  \centering
  \includegraphics[width=0.8\linewidth]{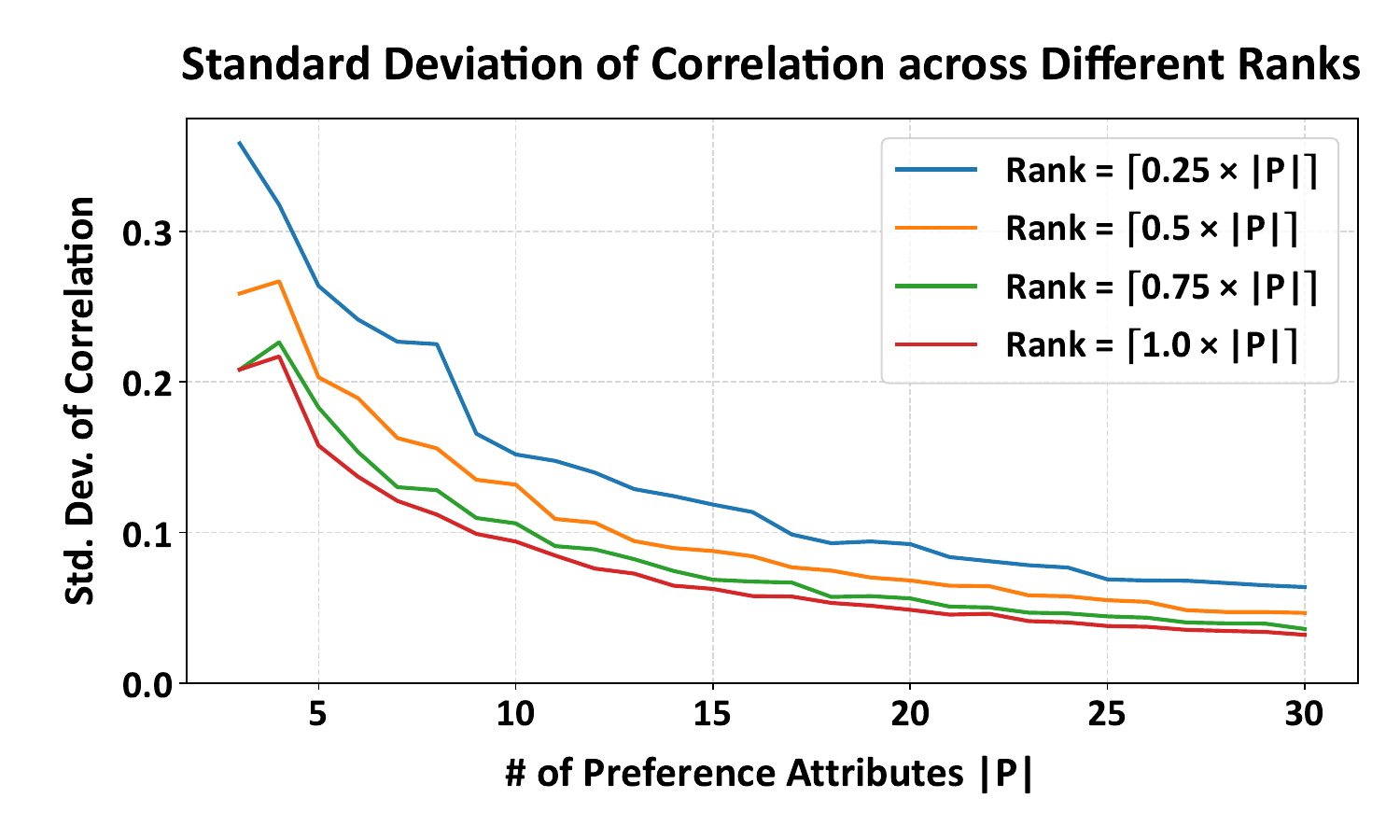}
  \vspace*{-0.05in}
  \caption{\replaceieeecr{}{
  Impact of the number of preference attributes on the standard deviation of per-variate correlation averages
  }}
  \label{fig:stdev_trend_varying_p}
\end{figure}
\begin{figure}[t]
  \centering
  \includegraphics[width=0.78\linewidth]{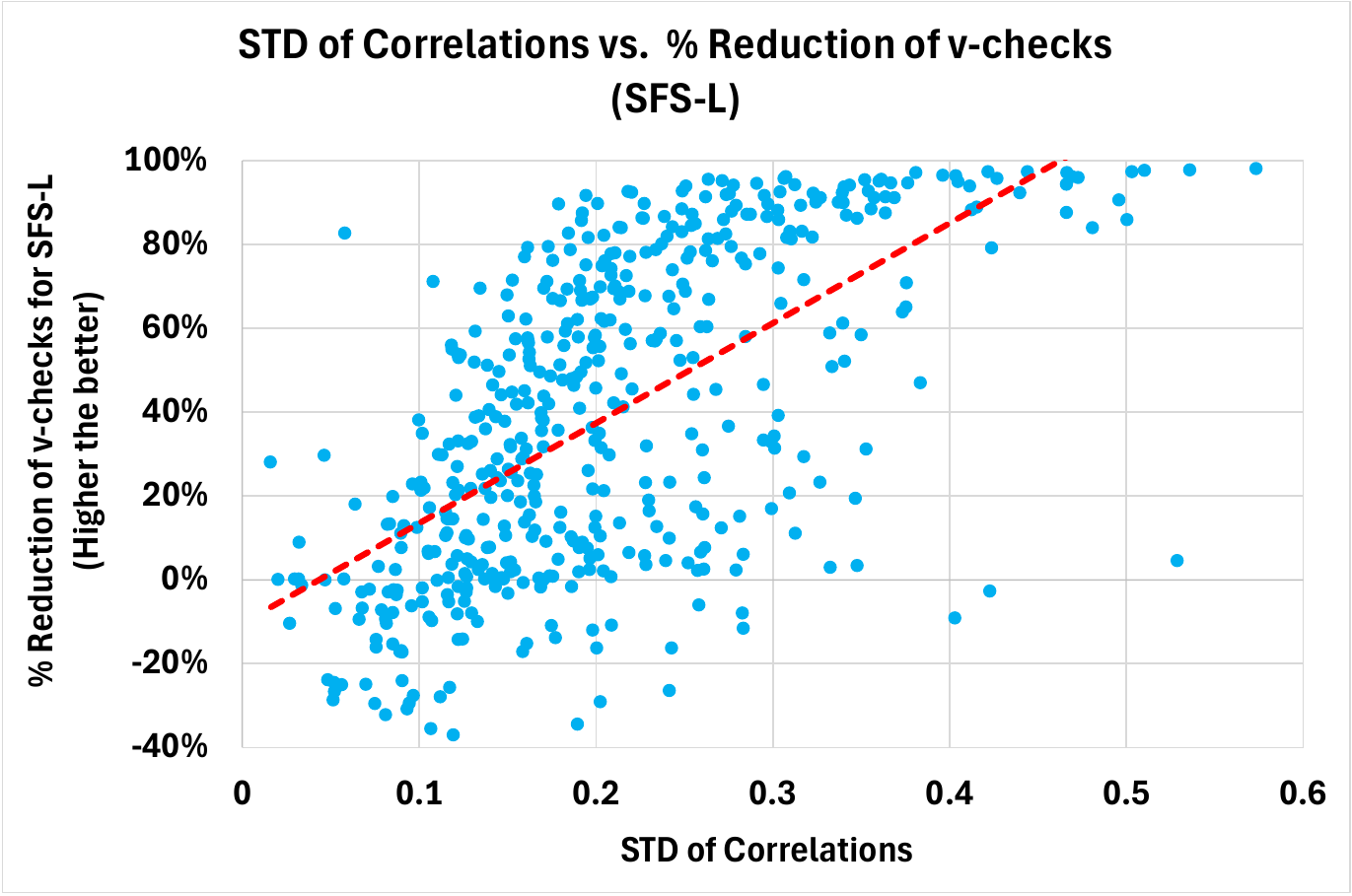}
  \caption{\replaceieeecr{}{
  Impact of the standard deviation of per-variate average correlations on the efficiency of SFS-L
  when using NCF variate ordering strategy \underline{(higher is better)} -- 100K data
  }}
  \label{fig:reduc_in_exec_time_varying_sdev}
  \vspace*{-0.05in}
\end{figure}

In Figure~\ref{fig:reduc_in_exec_time_varying_sdev}, we analyze how this reduction in standard deviation of per-variate correlation averages impacts the performance of
the SFS-L skyline algorithm when using the NCF
variate ordering
strategy.
As 
we see here, as would be expected,
the benefit of using the {\em min-correlation-first} (NCF) strategy increases with the increasing standard deviation (i.e., diversity) of the correlations. 
This is intuitive, 
as NCF places variates with lower average correlations earlier in the comparison order, where they are more likely to disprove tuple dominance. This allows 
{\em t-checks}
to terminate after comparing fewer variates, reducing the overall computation cost.
Importantly, when the standard deviations are low, we may see lower, or in fact in some cases negative, gains. 

We note that we can analyze the {\em direct} impact of the number of preference attributes on the gain by applying conditioning~\cite{Pearl2000, Pearl2009} on the data correlation, which would (as illustrated in Figure~\ref{fig:causal}(b)) eliminate the indirect causal paths impacting the gain.
We report these results in Table~\ref{tab:ablation_1_conditioned}. As we see in this table, when we condition by focusing on a relatively small range of correlations,  gain increases
for all algorithms as we increase the size of preference set, indicating that variate pruning opportunities become more significant as we add more preference attributes to the decision problem.

}

\subsubsection{\replaceieee{}{Impact of the Rank of the Preference Attribute Correlation Matrix\replaceieeecr{ Rank}{}}}\label{sec:impact_pref_attr_corr_rank}

\phantom{.}
In \replaceieee{Figure~\ref{fig:rank}}{Table~\ref{tab:rank}}, we see 
that the rank of the correlation matrix of preference attributes \replaceieeecr{have significant impact on}{does impact} the gains:   for the sort-based algorithms, SFS and SaLSa, gains are largest for low \replaceieee{or very high (full)}{ranks}; in contrast the divide-and-\replaceieee{concur}{conquer} based approach, D\&C, has the opposite behavior and sees the highest gains for mid-range ranks. These indicate that 
algorithms and  
VOS strategies are sensitive to the correlation structure.
\replaceieeecr{}{This is also in accordance with the trends we have observed in 
Figure~\ref{fig:stdev_trend_varying_p}: as we see in this figure, the curve describing the relationship between the number of preference attributes and the standard deviation of correlations flatten (lowering the overall diversity in correlations) as we increase the rank of the data and this explains why the gains tend to drop for data with large ranks. 
This implies that the proposed correlation based ranking strategies are especially useful for low rank data sets.}

\begin{table}[t]
  \caption{\replaceieee{}{The impact of lex sorting on the execution time (seconds); 
  (100K, synthetic data, $|P|$ = 5, v-check order is NCF on all cases; the table is per-row colored from best/green to worst/red)
  }}   \label{tab:lexsort}%
  \vspace*{-0.05in}
  \centering
  \renewcommand{\arraystretch}{1.2}
  \adjustbox{width=\columnwidth}{
    \begin{tabular}{|r|cccccc|}
    \hline
          & \multicolumn{6}{c|}{\textbf{SFS}} \\
    \cline{2-7}
    \multicolumn{1}{|r  |}{\textbf{Rank}} & \multicolumn{1}{P{3em}}{\textbf{NCF  LIFO}} & \multicolumn{1}{P{3em}}{\textbf{NCF  FIFO}} & \multicolumn{1}{P{3em}}{\textbf{XCF  LIFO}} & \multicolumn{1}{P{3em}}{\textbf{XCF  FIFO}} & \multicolumn{1}{P{4em}}{\textbf{Baseline} \replaceieeecr{}{$\langle sum\rangle$}} & \multicolumn{1}{P{6em}|}{\textbf{Baseline} \replaceieeecr{}{$\langle max, sum\rangle$}} \\
    \hline
    \textbf{3} & \cellcolor[rgb]{ 0,  .69,  .314}12.1 & \cellcolor[rgb]{ .937,  .737,  .737}17.3 & \cellcolor[rgb]{ .596,  .875,  .722}14.5 & \cellcolor[rgb]{ .753,  0,  0}20.4 & \cellcolor[rgb]{ .725,  .914,  .812}15.1 & \cellcolor[rgb]{ .804,  .196,  .196}19.6 \\
    \textbf{4} & \cellcolor[rgb]{ 0,  .69,  .314}2.8 & \cellcolor[rgb]{ .757,  .004,  .004}5.3 & \cellcolor[rgb]{ .525,  .851,  .675}3.7 & \cellcolor[rgb]{ .773,  .071,  .071}5.2 & \cellcolor[rgb]{ .553,  .859,  .69}3.7 & \cellcolor[rgb]{ .753,  0,  0}5.3 \\
    \textbf{5} & \cellcolor[rgb]{ 0,  .69,  .314}0.4 & \cellcolor[rgb]{ .753,  0,  0}1.2 & \cellcolor[rgb]{ .584,  .871,  .714}0.6 & \cellcolor[rgb]{ .933,  .725,  .725}0.9 & \cellcolor[rgb]{ .545,  .859,  .686}0.6 & \cellcolor[rgb]{ .878,  .506,  .506}1.0 \\
    \hline
    \end{tabular}%
  }

  \vspace{0.5em}
    \centering
  \adjustbox{width=\columnwidth}{
    \begin{tabular}{|r|cccccc|}
    \hline
          & \multicolumn{6}{c|}{\textbf{SaLSa}}\\
    \cline{2-7}
    \multicolumn{1}{|r  |}{\textbf{Rank}} & \multicolumn{1}{P{3em}}{\textbf{NCF  LIFO}} & \multicolumn{1}{P{3em}}{\textbf{NCF  FIFO}} & \multicolumn{1}{P{3em}}{\textbf{XCF  LIFO}} & \multicolumn{1}{P{3em}}{\textbf{XCF  FIFO}} & \multicolumn{1}{P{4em}}{\textbf{Baseline} \replaceieeecr{}{$\langle sum\rangle$}} & \multicolumn{1}{P{6em}|}{\textbf{Baseline} \replaceieeecr{}{$\langle max, sum\rangle$}} \\
    \hline
    \textbf{3} & \cellcolor[rgb]{ 0,  .69,  .314}12.2 & \cellcolor[rgb]{ .945,  .769,  .769}17.3 & \cellcolor[rgb]{ .655,  .89,  .765}14.9 & \cellcolor[rgb]{ .753,  0,  0}20.4 & \cellcolor[rgb]{ .769,  .925,  .843}15.4 & \cellcolor[rgb]{ .769,  .063,  .063}20.2 \\
    \textbf{4} & \cellcolor[rgb]{ 0,  .69,  .314}2.8 & \cellcolor[rgb]{ .796,  .165,  .165}5.3 & \cellcolor[rgb]{ .561,  .863,  .698}3.8 & \cellcolor[rgb]{ .808,  .22,  .22}5.2 & \cellcolor[rgb]{ .573,  .867,  .706}3.8 & \cellcolor[rgb]{ .753,  0,  0}5.4 \\
    \textbf{5} & \cellcolor[rgb]{ 0,  .69,  .314}0.4 & \cellcolor[rgb]{ .753,  0,  0}1.2 & \cellcolor[rgb]{ .608,  .878,  .729}0.6 & \cellcolor[rgb]{ .937,  .741,  .741}0.9 & \cellcolor[rgb]{ .573,  .867,  .706}0.6 & \cellcolor[rgb]{ .871,  .478,  .478}1.0 \\
    \hline
    \end{tabular}%
  }
\vspace*{-0.05in}
\end{table}%

\subsubsection{Impact of Variate-Order Aware Lexicographical Sorting}

In Table~\ref{tab:lexsort}, we present the execution times of various versions of the pre-sorting based algorithms: leveraging alternative lexicographical sorting strategies, along with baselines that rely on more conventional sorting orders.
As predicted in Section~\ref{sec:lvos}, lexicographical sorting with  {\em min-correlation-first} (NCF) variate-ordering and with LIFO skyline validation provides the best overall execution time performance. 

\subsubsection{Impact of Hardware Vectorization}


\replaceieee{Figure~\ref{fig:simd_results}}{Table~\ref{tab:simd_results}} shows the performance gains under scalar \replaceieeecr{as well as}{and} SIMD architectures. As we discussed in Section~\ref{Sec:simd}, in SIMD architectures, variate ordering helps through chunk pruning during dominance checks, rather than variate-pruning.
\replaceieeecr{As we see in this table, \replacenewa{}{for the SFS, SaLSa, and the D\&C algorithms,}}{For SFS, SaLSa and D\&C algorithms,} we see comparable gains in variate/chunk pruning and execution times with VOS
also when we rely on SIMD parallelism rather than performing  scalar computations. 

\section{Conclusions}
In this paper we argued that per-attribute dominance checks are a better proxy for skyline query execution time and that the number of these checks can be a function of the order in which the attributes are considered during skyline query processing.
Given this premise, we have shown that pairwise attribute correlations in the data can be a good signal for ordering the preference attributes and proposed several strategies for obtaining effective variate orders. 
We have further shown that 
a 
{\em min-correlation-first} attribute ordering strategy \replacenewa{to provide}{provides} significant gains in query execution times, especially for \replacenewa{large data sets with mixed pairwise attribute correlation portfolios}{\replaceieee{SalSa}{SaLSa} and D\&C algorithms and for larger number of variates}. 

\printbibliography

@book{Pearl2000,
  author    = {Judea Pearl},
  title     = {Causality: Models, Reasoning, and Inference},
  publisher = {Cambridge University Press},
  year      = {2000},
  edition   = {1st},
  isbn      = {978-0521773621}
}

@Article{Pearl2009,
  author    = {Pearl, Judea},
  journal   = {Statist. Surv.},
  title     = {Causal inference in statistics: An overview},
  year      = {2009},
  pages     = {96--146},
  volume    = {3},
  fjournal  = {Statistics Surveys},
  publisher = {The American Statistical Association, the Bernoulli Society, the Institute of Mathematical Statistics, and the Statistical Society of Canada},
}

@INPROCEEDINGS{distributedZorder, 
author={Jin Huang and Jian Chen and Qing Du and Jian Yin}, 
booktitle={FSKD}, title={A load balancing skyline query algorithm in high bandwidth distributed systems}, 
year={2010}, 
volume={5}, 
pages={2076 -2080}
}

@inproceedings{Zorder,
 author = {Lee, Ken C. K. and Zheng, Baihua and Li, Huajing and Lee, Wang-Chien},
 title = {Approaching the skyline in {Z} order},
 booktitle = {VLDB},
 year = {2007},
 pages = {279--290},
 numpages = {12}
}

@INPROCEEDINGS{Salsa,
  author = {Bartolini, Ilaria and Ciaccia, Paolo and Patella, Marco},
  title = {SaLSa: computing the skyline without scanning the whole sky},
  booktitle = {CIKM},
  year = {2006}
}

@INPROCEEDINGS{Borzsonyi01theskyline,
  author = {Stephan B{\"o}rzs{\"o}nyi and Donald Kossmann and Konrad Stocker},
  title = {The {Skyline} Operator},
  booktitle = {ICDE},
  year = {2001},
  pages = {421--430}
}

@INPROCEEDINGS{1129924,
  author = {Chaudhuri, Surajit and Dalvi, Nilesh and Kaushik, Raghav},
  title = {Robust Cardinality and Cost Estimation for {Skyline} Operator},
  booktitle = {ICDE},
  year = {2006},
  pages = {64}
}

@INPROCEEDINGS{Chomicki03skylinewith,
  author = {Jan Chomicki and Parke Godfrey and Jarek Gryz and Dongming Liang},
  title = {{Skyline} with presorting},
  booktitle = {ICDE},
  year = {2003},
  pages = {717--719}
}

@ARTICLE{KungLP75,
  author = {Hsiang-Tsung Kung and Fabrizio Luccio and Franco P. Preparata},
  title = {On Finding the Maxima of a Set of Vectors},
  journal = {J. ACM},
  year = {1975},
  volume = {22},
  pages = {469-476}
}

@INPROCEEDINGS{Kossmann02shootingstars,
  author = {Donald Kossmann and Frank Ramsak and Steffen Rost},
  title = {Shooting Stars in the Sky: {An} Online Algorithm for {Skyline} Queries},
  booktitle = {VLDB},
  year = {2002}
}

@ARTICLE{PapadiasTFS05,
  author = {Dimitris Papadias and Yufei Tao and Greg Fu and Bernhard Seeger},
  title = {Progressive {Skyline} computation in database systems},
  journal = {ACM Trans. Database Systems},
  year = {2005},
  volume = {30},
  pages = {41-82}
}

@ARTICLE{StojmenovicM88,
  author = {Ivan Stojmenovi{\'c} and Masahiro Miyakawa},
  title = {An optimal parallel algorithm for solving the maximal elements problem
	in the plane},
  journal = {Parallel Computing},
  year = {1988},
  volume = {7},
  pages = {249-251}
}

@INPROCEEDINGS{Wu06parallelizingskyline,
  author = {Ping Wu and Caijie Zhang and Ying Feng and Ben Y. Zhao and Divyakant
	Agrawal and Amr El Abbadi},
  title = {Parallelizing {Skyline} queries for scalable distribution},
  booktitle = {EDBT},
  year = {2006},
  pages = {112--130}
}

@inproceedings{needle_in_haystack,
  author={Schlake, Georg Stefan and Beecks, Christian},
  booktitle={2024 IEEE International Conference on Big Data (BigData)}, 
  title={The Skyline Operator to Find the Needle in the Haystack for Automated Clustering}, 
  year={2024},
  volume={},
  number={},
  pages={6117-6122},
  doi={10.1109/BigData62323.2024.10825416}}

@article{skyline_on_massive_data,
  title={Computing Prominent Skyline on Massive Data},
  author={Wan, Xiaolong and Han, Xixian and Wang, Jinbao},
  journal={Data Science and Engineering},
  pages={1--30},
  year={2024},
  publisher={Springer}
}

@inproceedings{godfrey2005maximal,
  title={Maximal vector computation in large data sets},
  author={Godfrey, Parke and Shipley, Ryan and Gryz, Jarek and others},
  booktitle={VLDB},
  volume={5},
  pages={229--240},
  year={2005}
}

@article{bentley1993fast,
  title={Fast linear expected-time algorithms for computing maxima and convex hulls},
  author={Bentley, Jon L and Clarkson, Kenneth L and Levine, David B},
  journal={Algorithmica},
  volume={9},
  pages={168--183},
  year={1993},
  publisher={Springer}
}

@inproceedings{park2009parallel,
  title={Parallel skyline computation on multicore architectures},
  author={Park, Sungwoo and Kim, Taekyung and Park, Jonghyun and Kim, Jinha and Im, Hyeonseung},
  booktitle={2009 IEEE 25th International Conference on Data Engineering},
  pages={760--771},
  year={2009},
  organization={IEEE}
}

@article{beinecke2021gaussian,
  title={Gaussian noise up-sampling is better suited than SMOTE and ADASYN for clinical decision making},
  author={Beinecke, Jacqueline and Heider, Dominik},
  journal={BioData Mining},
  volume={14},
  number={1},
  pages={49},
  year={2021},
  publisher={Springer}
}

@misc{Hindy_2021,
  title={Hong Kong Weather (1970-2019)},
  url={https://www.kaggle.com/datasets/hindy51/hong-kong-weather-20082016/data},
  journal={Kaggle},
  author={Hindy},
  year={2021},
  month={Jan}
}

@incollection{keahey2020lessons,
  title={Lessons Learned from the Chameleon Testbed},
  author={Kate Keahey and Jason Anderson and Zhuo Zhen and Pierre Riteau and Paul Ruth and Dan Stanzione and Mert Cevik and Jacob Colleran and Haryadi S. Gunawi and Cody Hammock and Joe Mambretti and Alexander Barnes and Fran\c{c}ois Halbach and Alex Rocha and Joe Stubbs},
  booktitle={Proceedings of the 2020 USENIX Annual Technical Conference (USENIX ATC '20)},
  publisher={USENIX Association},
  month={July},
  year={2020}
}

@article{zhang2014efficient,
  title={An efficient approach to nondominated sorting for evolutionary multiobjective optimization},
  author={Zhang, Xingyi and Tian, Ye and Cheng, Ran and Jin, Yaochu},
  journal={IEEE Transactions on Evolutionary Computation},
  volume={19},
  number={2},
  pages={201--213},
  year={2014},
  publisher={IEEE}
}

@article{ZHANG2026130889,
title = {Skyline operators in multi-criteria decision making: A review of characterization, comparison, and perspectives},
journal = {Expert Systems with Applications},
volume = {306},
%pages = {130889},
year = {2026},
issn = {0957-4174},
doi = {https://doi.org/10.1016/j.eswa.2025.130889},
url = {https://www.sciencedirect.com/science/article/pii/S095741742504504X},
author = {Xichen Zhang and Hasan Cavusoglu},
keywords = {Multi-criteria decision-making, Skyline, Decision support system}
}

@INPROCEEDINGS{7373349,
  author={Wen, Yu-Ting and Cho, Kae-Jer and Peng, Wen-Chih and Yeo, Jinyoung and Hwang, Seung-won},
  booktitle={2015 IEEE International Conference on Data Mining}, 
  title={KSTR: Keyword-Aware Skyline Travel Route Recommendation}, 
  year={2015},
  volume={},
  number={},
  pages={449-458},
  doi={10.1109/ICDM.2015.37}}

@inproceedings{10.1145/1559845.1559899,
author = {Zhang, Zhenjie and Yang, Yin and Cai, Ruichu and Papadias, Dimitris and Tung, Anthony},
title = {Kernel-based skyline cardinality estimation},
year = {2009},
isbn = {9781605585512},
publisher = {Association for Computing Machinery},
address = {New York, NY, USA},
url = {https://doi-org.ezproxy1.lib.asu.edu/10.1145/1559845.1559899},
doi = {10.1145/1559845.1559899},
booktitle = {Proceedings of the 2009 ACM SIGMOD International Conference on Management of Data},
pages = {509–522},
numpages = {14},
location = {Providence, Rhode Island, USA},
series = {SIGMOD '09}
}

@InProceedings{10.1007/978-3-540-24627-5_7,
author="Godfrey, Parke",
editor="Seipel, Dietmar
and Turull-Torres, Jos{\'e} Mar{\'i}a",
title="Skyline Cardinality for Relational Processing",
booktitle="Foundations of Information and Knowledge Systems",
year="2004",
publisher="Springer Berlin Heidelberg",
address="Berlin, Heidelberg",
pages="78--97",
isbn="978-3-540-24627-5"
}

@article{bentley1978average,
  title={On the average number of maxima in a set of vectors and applications},
  author={Bentley, Jon Louis and Kung, Hsiang-Tsung and Schkolnick, Mario and Thompson, Clark D},
  journal={Journal of the ACM (JACM)},
  volume={25},
  number={4},
  pages={536--543},
  year={1978},
  publisher={ACM New York, NY, USA}
}

@article{10.1145/3588958,
author = {Miao, Xiaoye and Wu, Yangyang and Peng, Jiazhen and Gao, Yunjun and Yin, Jianwei},
title = {Efficient and Effective Cardinality Estimation for Skyline Family},
year = {2023},
issue_date = {May 2023},
publisher = {Association for Computing Machinery},
address = {New York, NY, USA},
volume = {1},
number = {1},
url = {https://doi-org.ezproxy1.lib.asu.edu/10.1145/3588958},
doi = {10.1145/3588958},
journal = {Proc. ACM Manag. Data},
month = may,
articleno = {104},
numpages = {21}
}

@article{luo2012sampling,
  title={A sampling approach for skyline query cardinality estimation},
  author={Luo, Cheng and Jiang, Zhewei and Hou, Wen-Chi and He, Shan and Zhu, Qiang},
  journal={Knowledge and information systems},
  volume={32},
  number={2},
  pages={281--301},
  year={2012},
  publisher={Springer}
}

@article{10.1145/1893173.1893176,
author = {Cho, Sung-Ryoung and Lee, Jongwuk and Hwang, Seung-Won and Han, Hwansoo and Lee, Sang-Won},
title = {VSkyline: vectorization for efficient skyline computation},
year = {2010},
issue_date = {June 2010},
publisher = {Association for Computing Machinery},
address = {New York, NY, USA},
volume = {39},
number = {2},
issn = {0163-5808},
url = {https://doi.org/10.1145/1893173.1893176},
doi = {10.1145/1893173.1893176},
journal = {SIGMOD Rec.},
month = dec,
pages = {19–26},
numpages = {8}
}

@INPROCEEDINGS{582015,
  author={Leutenegger, S.T. and Lopez, M.A. and Edgington, J.},
  booktitle={Proceedings 13th International Conference on Data Engineering}, 
  title={STR: a simple and efficient algorithm for R-tree packing}, 
  year={1997},
  volume={},
  number={},
  pages={497-506},
  doi={10.1109/ICDE.1997.582015}
}

@article{10.1145/3802026,
author = {Mandal, Pratanu and Gorantla, Abhinav and Candan, K. Sel\c{c}uk and Sapino, Maria Luisa},
title = {Causal Search for Skylines (CSS): Causally-Informed Selective Data De-Correlation},
year = {2026},
issue_date = {June 2026},
publisher = {Association for Computing Machinery},
address = {New York, NY, USA},
volume = {4},
number = {3},
url = {https://doi.org/10.1145/3802026},
doi = {10.1145/3802026},
journal = {Proc. ACM Manag. Data},
month = may,
articleno = {149},
numpages = {27}
}

@inproceedings{causalbench_er,
author = {Kapki{\c c}, Ahmet and Mandal, Pratanu and Gorantla, Abhinav and Wan, Shu and {\c C}oban, Ertu{\u g}rul and Sheth, Paras and Liu, Huan and Candan, K. Sel{\c c}uk},
title = {CausalBench-ER: Causally-Informed Explanations and Recommendations for Reproducible Benchmarking},
year = {2025},
isbn = {9798400720406},
publisher = {Association for Computing Machinery},
address = {New York, NY, USA},
url = {https://doi.org/10.1145/3746252.3761606},
doi = {10.1145/3746252.3761606},
booktitle = {Proceedings of the 34th ACM International Conference on Information and Knowledge Management},
pages = {6426–6431},
numpages = {6},
location = {Seoul, Republic of Korea},
series = {CIKM '25}
}

@online{causalbenchgrant,
author={Candan, K. Selcuk and Liu, Huan},
title={NSF OAC Grant \# 2311716: Elements: CausalBench: A Cyberinfrastructure for Causal-Learning Benchmarking for Efficacy, Reproducibility, and Scientific Collaboration},
url={https://www.nsf.gov/awardsearch/showAward?AWD_ID=2311716},
urldate={2025-08-14},
journal={NSF AWARD SEARCH: Award \# 2311716},
year={2023},
month ={jun}
}

@misc{uci_ml_repo,
  author       = {Kelly, Markelle and Longjohn, Rachel and Nottingham, Kolby},
  title        = {The {UCI} Machine Learning Repository},
  howpublished = {\url{https://archive.ics.uci.edu}},
  url          = {https://archive.ics.uci.edu}
}

@misc{gorantla2026vosvariateorderingstrategies,
      title={VOS: Variate Ordering Strategies for Skyline Query Optimization}, 
      author={Abhinav Gorantla and Pratanu Mandal and K. Selçuk Candan and Maria Luisa Sapino},
      year={2026},
      eprint={2608.26464},
      archivePrefix={arXiv},
      primaryClass={cs.DB},
      url={https://arxiv.org/abs/2608.26464v2}, 
      note   = {arXiv:2608.26464v2, \url{https://arxiv.org/abs/2608.26464v2}},
}

\end{document}